\documentclass[final]{IEEEtran}
\usepackage{amsthm,amssymb,graphicx,multirow,amsmath,color,amsfonts}
\usepackage[update,prepend]{epstopdf}
\usepackage[noadjust]{cite}
\usepackage[latin1]{inputenc}
\usepackage{tikz}
\usepackage{bm}
\usepackage{bbm}  
\usepackage[nolist]{acronym}
\usepackage{pdfpages}
\usepackage{multirow}
\usepackage{subfig}
\usepackage{comment}
\usepackage{fancyhdr}
\usepackage{booktabs}

\def\nb0{{\mathbf{0}}}
\def\nb1{{\mathbf{1}}}

\newtheorem{prop}{Proposition}

\newtheorem{remark}{Remark}

\allowdisplaybreaks 
\usepackage{algorithm}
\usepackage{algorithmic}
\usepackage{setspace}	

\begin{acronym}

\acro{BS}{base station}

\acro{LoS}{line-of-sight}
\acro{NLoS}{non-line-of-sight}
\acro{UPA}{uniform planar array}
\acro{VALSE}{VAriational Line Spectral Estimation}
\acro{EM}{expectation-maximization}
\acro{RANSAC}{random sample consensus}
\acro{WLS}{weighted least squares}
\acro{LS}{least squares}

\acro{CRB}{Cram{\'e}r-Rao bound}
\acro{CRLB}{Cram{\'e}r-Rao lower bound}
\acro{BCRB}{Bayesian Cram{\'e}r-Rao bound}
\acro{cs}[CS]{compressed sensing}   

\acro{RMSE}{root mean-square error}    
\acro{NMSE}{normalized mean-square error}

\acro{w.r.t.}{with respect to}
\acro{ELAA}{extremely large antenna array}
\acro{CSF}{Chirp-coupled Subarray Far-field}
\acro{CHARM}{CHirp-coupled Angular-Range estiMation}
\acro{MUSIC}{MUltiple SIgnal
Classification}

\acro{ULA}{uniform linear array}
\acro{UPA}{uniform planar array}
\acro{OMP}{Orthogonal Matching Pursuit}

\acro{ML}{maximum likelihood}

\acro{PDF}{probability density function}

\acro{SNR}{signal-to-noise ratio}

\end{acronym}

\begin{document}
\graphicspath{{./figures/}}
\title{
Scalable High-Precision Near-Field Channel Parameter Estimation via Spatial Chirp Structure  
}
\author{Lin Chen, \IEEEmembership{Graduate Student Member, IEEE}, Xiaojun Yuan, {\em Fellow, IEEE}, \\and Ying-Jun Angela Zhang, {\em Fellow, IEEE}
		\thanks{ 
			Lin Chen and Ying-Jun Angela Zhang are with the Department of Information Engineering, The Chinese University of Hong Kong (CUHK), Hong Kong (e-mail: lin.chen@link.cuhk.edu.hk,~yjzhang@ie.cuhk.edu.hk).
			
			Xiaojun Yuan is with the National Key Laboratory of Wireless Communications, University of Electronic Science and Technology of China (UESTC), Chengdu 611731, China (e-mail: xjyuan@uestc.edu.cn).
		} 
\vspace{-4mm}}

\maketitle

\begin{abstract}

This paper presents a scalable framework for high-precision near-field multipath channel parameter estimation in extremely large antenna array (ELAA) systems,
enabling joint recovery of path number, path gains, angles, and ranges from a single noisy observation.  
The key idea is to interpret the near-field multipath channel as a superposition of spatial chirp components with spatially varying frequencies and exploit this structure through a partitioned ELAA architecture. Specifically, we establish a Chirp-coupled Subarray Far-field (CSF) model, where each near-field path is locally represented as a far-field sinusoid with a constant spatial frequency within each subarray, while these local spatial frequencies are coupled across subarrays through a linear relationship induced by the underlying spatial chirp, forming a path-specific chirp trajectory.
Based on this model, we propose the CHirp-coupled Angular-Range estiMation (CHARM) algorithm, which performs gridless local frequency estimation followed by cross-subarray trajectory recovery. 
To mitigate the potential modeling mismatch of the CSF model, we further propose the enhanced CHARM (E-CHARM) algorithm, which refines the CHARM estimate under the near-field channel model through maximum likelihood. The computational complexity of the proposed algorithms scales linearly with the array size. Moreover, simulation results show that the proposed algorithms achieve reliable path-number detection, high-precision angle--range estimation, and accurate channel reconstruction.
\end{abstract}


\acresetall

\vspace{-5mm}

\section{Introduction}

\Ac{ELAA} is expected to provide unprecedented array gain and spatial resolution for high-quality communication and sensing, supporting emerging 6G applications such as autonomous driving, low-altitude economy, and smart cities~\cite{6G_freq_antenna,6G_freq}.
The resulting large-array aperture significantly extends the radiative near-field region to typical communication ranges, where the propagation is governed by spherical rather than planar wavefronts~\cite{NFtutorial_Dai,selvan2017fraunhofer}. 
Unlike the {\em far-field planar} wavefront, which induces a {\em linear} phase profile across the array and is characterized primarily by its propagation angle, the {\em near-field spherical} wavefront exhibits an angle- and range-dependent phase profile that is approximately {\em quadratic} across the array. This additional {\em range} dimension enables spatial focusing~\cite{chen2025quasi}, user localization~\cite{yuan2024scalable}, user tracking~\cite{chen2026tracking}, and environment sensing~\cite{zhou2024single}, and at the same time increases the difficulty  of channel acquisition~\cite{NFtutorial}.

For an ELAA with hundreds or even thousands of antennas, direct estimation of the channel coefficients is inherently high-dimensional and computationally challenging.
However, practical propagation is typically sparse, consisting of a limited number of user- and scatterer-induced paths~\cite{NFtutorial}.
This sparse multipath structure enables a low-dimensional parametric representation of the near-field channel, characterized by the path number (i.e., model order), the complex gain of each path, and the corresponding scatterer/user positions (parameterized by their angle and range)~\cite{NFCE1,NFCE2}.   
This motivates {\em parametric channel estimation} as an alternative to direct per-antenna channel estimation, where the objective is to infer these channel parameters from limited observations~\cite{kosasih2023parametric,NFCE1,NFCE2}.
Once estimated, these parameters enable full near-field channel reconstruction and, more importantly, provide explicit {\em geometric information} for beam focusing, user localization, and environment perception.

Existing near-field parameter estimation methods, however, face two major challenges: \emph{computational scalability} and \emph{estimation accuracy}, particularly when only a \emph{single observation} is available.
Subspace-based methods, such as \ac{MUSIC}, can estimate angle--range parameters by identifying peaks in a near-field spatial spectrum constructed from the signal subspace~\cite{MUSIC0,kosasih2023parametric,MUSIC2}. Since the signal subspace is obtained from the received covariance matrix, reliable estimation generally requires {\em multiple} observations, while the associated matrix decomposition incurs {\em cubic-order} complexity in the array size.
Sparse-recovery approaches, such as \ac{OMP}, 
represent the channel over a discretized angle--range dictionary and identify the active atoms in the dictionary from the received signal~\cite{cui2022channel,xu2025near}, enabling single-observation estimation.
However, the two-dimensional angle--range grid leads to a large dictionary and substantial storage overhead. Such approaches generally exhibit {\em quadratic-order} computational complexity \ac{w.r.t.} the array size~\cite{cui2022channel}, while parameter {\em discretization} inherently limits the estimation accuracy.
Off-grid refinement methods improve the resolution of on-grid estimates by iteratively refining the angle--range parameters in the continuous domain~\cite{cui2022channel,offgrid1}, but their estimation accuracy remains sensitive to the grid-based {\em initialization}. 
Semi-gridless and gridless methods in~\cite{semi-grid,gridless-chirp} further improve the parameter estimation accuracy, while their computational complexity still exhibits  {\em quadratic}~\cite{semi-grid} or {\em even higher-order}~\cite{gridless-chirp} scaling with the array size, limiting their scalability to ELAAs.
Moreover, many existing methods assume that the path number is known {\em a priori}, which further limits their applicability to practical multipath channel estimation.
Therefore, reliably recovering the unknown path number and estimating the angle--range parameters with high precision from a single noisy observation, while maintaining computational complexity scalable to the ELAA size, remains challenging.

To tackle the above challenges, we first seek to reduce the computational burden through \emph{array partitioning}. Rather than directly processing the high-dimensional full-array near-field observation, we partition the ELAA into multiple sufficiently small subarrays and perform low-dimensional inference locally.
Interestingly, although a user or scatterer may lie in the near-field region of the entire ELAA, it can be approximately regarded as a far-field source relative to each sufficiently small subarray~\cite{PWFF}.
Consequently, for each propagation path, the global  \emph{quadratic} phase variation across the full ELAA becomes approximately \emph{linear} within each subarray.
The local subarray observation of the multipath channel therefore follows a sum-of-sinusoids model, where each sinusoidal component is characterized by a local spatial frequency corresponding to the path angle observed at that subarray.
This enables existing gridless line spectral estimation methods, such as \ac{VALSE}~\cite{VALSE}, to jointly infer the local spatial frequencies, path number, and path gains from a single subarray observation.
 
To efficiently recover the global near-field angle--range parameters by fusing these local estimates, 
we reveal a key cross-subarray structure: the local spatial frequencies associated with the same physical path are not arbitrary, but exhibit a \emph{linear variation} across subarrays. This structure originates from the inherent \emph{spatial-chirp structure} of near-field propagation.
Specifically, over the entire ELAA, the spherical wavefront induces a \emph{quadratic} phase variation with the antenna index. Such a quadratic phase profile can be interpreted as a spatial chirp whose spatial frequency varies linearly across antennas and is jointly determined by the path angle and range.
Hence, the local frequencies associated with each physical path align along a straight line across subarrays, providing a structured relationship for recovering the global angle and range.

In multipath channels, however, exploiting this linear relationship is nontrivial, since the local components are estimated independently at different subarrays and their cross-subarray path correspondences are unknown.
Moreover, noise and limited inter-path separability may lead to missed or spurious local components, further complicating the recovery of the underlying global paths. In addition, the local far-field approximation inevitably introduces modeling mismatch that may limit the ultimate estimation accuracy.
We therefore develop practical algorithms that exploit the above {\em gridless local-inference} and {\em structured global-fusion} framework while explicitly addressing these imperfections, enabling scalable high-precision near-field channel parameter estimation. The main contributions are summarized below.

\begin{itemize}

\item  
We establish a \ac{CSF} model. Under this model, each propagation path corresponds to a spatial chirp component, whose {\em initial frequency} and {\em chirp rate} are determined by the angle--range parameters. Each chirp component is locally represented as a sinusoid, while its local spatial frequencies vary {\em linearly} across subarrays according to the chirp rate, forming a path-specific {\em chirp trajectory}.
This reformulates near-field channel parameter estimation as subarray-level line-spectral inference followed by cross-subarray trajectory recovery.

\item Building on this reformulation, we propose the \ac{CHARM} algorithm. 
We first apply \ac{VALSE}~\cite{VALSE} to each subarray to obtain local estimates together with their uncertainties. To address the unordered and imperfect nature of local estimates, we formulate chirp trajectory recovery as a probabilistic association and parameter estimation problem. 
The resulting \ac{EM} framework jointly infers the data associations, chirp trajectories, and path number.
CHARM thus enables gridless angle--range recovery and channel reconstruction 
with computational complexity that {\em scales linearly} with the array size.

\item  
To mitigate the potential {\em modeling mismatch} of the CSF approximation, 
we further develop the enhanced CHARM (E-CHARM) algorithm. Using CHARM as a {\em reliable initialization}, E-CHARM refines the channel parameters under the near-field model according to the \ac{ML} principle.  
E-CHARM achieves this refinement with a moderate computational overhead that {\em scales linearly} with the array size. 

\end{itemize}

\vspace{-2mm}
 
\subsection{Related Works}
 
The quadratic phase variation of a near-field array response has long been interpreted as a spatial chirp~\cite{1988NFchirp}.
This structure has been exploited for near-field source localization~\cite{1988NFchirp,2018NFchirp} and near-field beam training~\cite{2023NFchirp}, and more recently, for channel parameter estimation~\cite{gridless-chirp}.
However, direct full-array near-field processing leads to high computational complexity, e.g., the method in~\cite{gridless-chirp} scales cubically \ac{w.r.t.} the array size. 
In contrast, our work exploits the spatial-chirp structure through the linear variation of local spatial frequencies across subarrays.
This allows the near-field channel parameter estimation from a high-dimensional full-array observation to be reformulated as low-dimensional local inference from subarray observations, followed by structured cross-subarray fusion, thereby achieving computational complexity scalable to the ELAA size.

Array partitioning has also been adopted in the literature for near-field channel modeling and parameter estimation.
It allows the full-array near-field response to be represented by multiple subarray far-field models associated with subarray-dependent angles~\cite{PWFF,yuan2024scalable}.
Based on this representation, the authors in \cite{yuan2024scalable} first estimate the local angles observed by different subarrays and then fuse them through their nonlinear geometric relationship with the source position. 
However, only a single propagation path is considered, such that the local angle estimates obtained at different subarrays naturally correspond to the same physical path. In multipath channels, the correspondences among the local estimates become unknown, introducing an additional path-association problem.
The proposed CSF model explicitly characterizes the linear relationship among the local spatial frequencies across subarrays, enabling efficient cross-subarray fusion with joint path association and global parameter recovery.

 \vspace{-2mm}
 
\subsection{Organization and Notation}
    The rest of this paper is organized as follows. Sec.~\ref{sec:sys} introduces the near-field multipath channel model and formulates the channel parameter estimation problem. Sec.~\ref{sec:model} establishes the CSF model and reformulates the problem. Sec.~\ref{sec:CHARM} develops the CHARM algorithm under the CSF model. Sec.~\ref{sec:E-CHARM} further develops the E-CHARM algorithm. 
    Sec.~\ref{sec:simu} discusses the simulation results. Sec.~\ref{sec:conclusion} concludes this paper.
		
		{\em Notation:} 
		The upper- and lower-case bold letters denote matrices and vectors, respectively. 
		For a matrix, $\mathbf{(\cdot)}^{\mathsf T}$, $\mathbf{(\cdot)}^{*}$, and $\mathbf{(\cdot)}^{\mathsf H}$ represent transpose, conjugate, and conjugate transpose operators, respectively. $\|\cdot\|_2$ denotes the $l_2$ norm. $\mathbf{I}$ denotes the identity matrix of an appropriate size. The imaginary unit is denoted by ${\rm j}$ with ${\rm j}^2=-1$.
        $\mathcal U(x;a,b)$ denotes the uniform \ac{PDF} with $\mathcal U(x;a,b)=\frac{1}{b-a}$ for $a\le x \le b$ and $0$ otherwise. Moreover, $\mathcal N(x;\mu,\sigma^2)$ denotes the Gaussian \ac{PDF} evaluated at $x$ with mean $\mu$ and variance $\sigma^2$, while $\mathcal{CN}(z;\mu,\sigma^2)$ denotes the circularly symmetric complex Gaussian \ac{PDF} evaluated at $z$ with mean $\mu$ and variance $\sigma^2$.

\vspace{-1mm}

 \section{System Model and Problem Description}\label{sec:sys}

\begin{figure}
    \centering
    \includegraphics[width=0.75\linewidth]{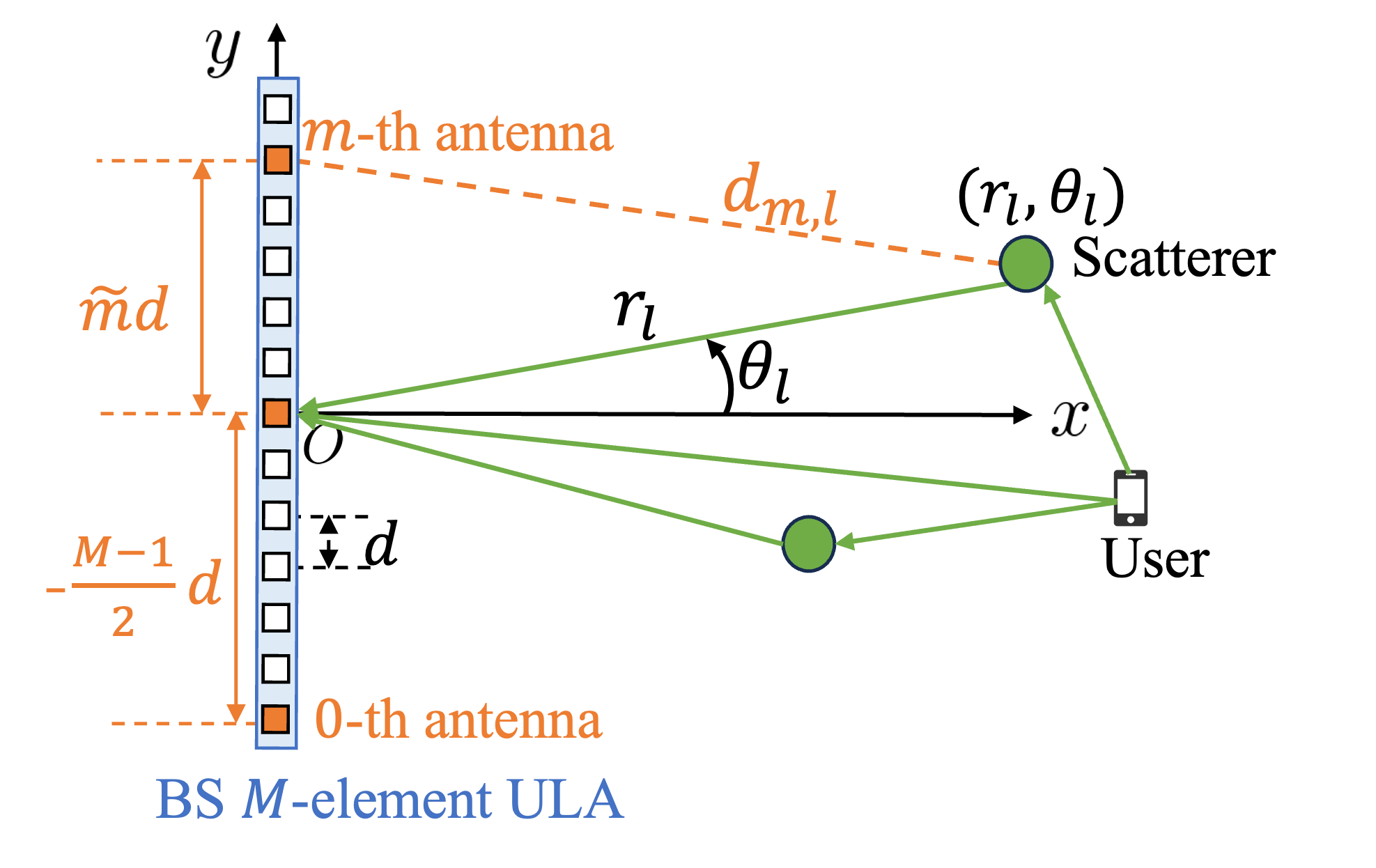}
    \vspace{-2mm}
    \caption{The uplink channel estimation scenario with the user and scatterers located in the near-field region of the BS array.}
    \vspace{-3mm}
    \label{fig:sys}
\end{figure}

We consider an uplink channel estimation between a multi-antenna \ac{BS} and a single-antenna user in a narrowband system, as shown in Fig.~\ref{fig:sys}. The channel is modeled as a superposition of multiple propagation paths.
The system is assumed to be operated at a carrier frequency $f_{\rm c}$ with wavelength $\lambda_{\rm c}$.
The BS is equipped with an \ac{ELAA} configured as a \ac{ULA} with $M$ antennas.\footnote{The proposed framework can be generalized to a \ac{UPA} using the corresponding two-dimensional near-field model.}
The antenna spacing is $d=\frac{\lambda_{\rm c}}{2}$, and the array aperture is $D=(M-1)d$. We establish a Cartesian coordinate system $\rm xOy$ with the origin $O$ at the BS's array center.
We focus on the radiative near-field regime, corresponding to the distance interval $[r_{\rm Fresnel}, r_{\rm Rayleigh}]$, where spherical-wave propagation effects cannot be neglected.
Here,
$r_{\rm Fresnel}=0.62\sqrt{\frac{D^3}{\lambda_{\rm c}}}$
and
$r_{\rm Rayleigh}=\frac{2D^2}{\lambda_{\rm c}}$
denote the Fresnel and Rayleigh distances, respectively~\cite{selvan2017fraunhofer}.
The user transmits one pilot signal, and the \ac{BS} utilizes the received single snapshot for the near-field multipath channel estimation.

\vspace{-2mm}

\subsection{Received Signal Model}

Let $x_{\rm p}$ denote the uplink pilot with $|x_{\rm p}|=1$. The received one-snapshot signal at the BS can be expressed as
\begin{align}
    \mathbf{y}_{\rm p}=\mathbf{h}x_{\rm p}+\mathbf{n}_{\rm p} \in \mathbb{C}^{M\times1},
\end{align}
where $\mathbf{h}\in\mathbb{C}^{M\times1}$ is the uplink channel, and $\mathbf{n}_{\rm p}\sim\mathcal{CN}(\mathbf{0},\sigma_{\rm n}^2\mathbf{I})$ is the measurement noise with power $\sigma_{\rm n}^2$.
Since the pilot is known at the receiver, we normalize the received signal as
\begin{align}\label{eq:y}
    \mathbf{y}=\frac{\mathbf{y}_{\rm p}}{x_{\rm p}}=\mathbf{h}+\mathbf{n},
\end{align}
where $\mathbf{n}=[n_0,\dots,n_m,\dots,n_{M-1}]^\mathsf{T}$ and $\mathbf{n}\sim\mathcal{CN}(\mathbf{0},\sigma_{\rm n}^2\mathbf{I})$.

Suppose the signal from the user to the BS experiences $L$ distinct propagation paths, as shown in Fig.~\ref{fig:sys}.
For generality, we use the term {\em source point} to denote the physical point associated with a propagation path, which corresponds to the user for a \ac{LoS} path and to a scatterer for a \ac{NLoS} path.   
The $\ell$-th source point is located at 
$[r_\ell\cos\theta_\ell,r_\ell\sin\theta_\ell]^{\mathsf{T}}$,  where $\theta_\ell$ and $r_\ell$ denote its angle and distance relative to the BS, respectively, and $\ell\in\{1,...,L\}$. {Moreover, we assume $\theta_\ell\in[\theta_{\min},\theta_{\max}]\subset(-\frac{\pi}{2},\frac{\pi}{2})$ and $r_\ell\in[r_{\min},r_{\max}]\subset (r_{\rm Fresnel},r_{\rm Rayleigh})$.}
Denote the antenna index by $m\in\{0,...,M-1\}$. 
The $m$-th antenna element is located at $[0,\tilde{m} d]^\mathsf{T}$, where $\tilde{m}=m-\frac{M-1}{2}$. 
The distance between the $m$-th antenna element and the $\ell$-th source point is given by
\begin{equation}\label{eq:dist}
d_{m,\ell}
=
\sqrt{
r_\ell^2 + (\tilde{m} d)^2 - 2 r_\ell \tilde{m} d \sin\theta_\ell
}.
\end{equation} 
The near-field channel vector $\mathbf{h}$ is the superposition of multipath components and can be expressed as~\cite{cui2022channel} 
\begin{align}\label{eq:h-exact}
    \mathbf{h}=\sum_{\ell=1}^{L}\alpha_\ell \mathbf{b}(\theta_\ell,r_\ell).
\end{align}
Here, $\alpha_\ell$ is the complex path gain of the $\ell$-th path, and $\mathbf{b}(\theta_\ell,r_\ell)$ is the near-field array response vector given by
\begin{align}\label{eq:b-exact}
    \mathbf{b}(\theta_\ell,r_\ell)=[e^{{\rm j}  \psi_{0,\ell}},...,e^{{\rm j} \psi_{m,\ell}},..., e^{{\rm j} \psi_{M-1,\ell}} ]^\mathsf{T},
\end{align}
where $\psi_{m,\ell}=-\frac{2\pi}{\lambda_{\rm c}} (d_{m,\ell}-r_\ell)$ 
denotes the phase difference of the received signal at the $m$-th antenna element relative to the array center for the $\ell$-th path. 
Based on the above near-field model, the received signal in \eqref{eq:y} can be rewritten as
\begin{align}\label{eq:y-matrix}
    \mathbf{y}&=\sum_{\ell=1}^{L}\alpha_\ell \mathbf{b}(\theta_\ell,r_\ell)+\mathbf{n} =\mathbf{B}(\bm\theta,\mathbf{r})\bm\alpha +\mathbf{n},
\end{align}
where $\bm\alpha=[\alpha_1,\dots,\alpha_L]^\mathsf T$, $\bm\theta=[\theta_1,\dots,\theta_L]^\mathsf T$, $\mathbf r=[r_1,\dots,r_L]^\mathsf T$, and $\mathbf B(\bm\theta,\mathbf r)
=
\left[
\mathbf b(\theta_1,r_1),
\dots,
\mathbf b(\theta_L,r_L)
\right]$. 

\vspace{-2mm}

\subsection{Near-Field Spatial Chirp Model}  

In the radiative near-field regime, the distance $d_{m,\ell}$ can be approximated with its second-order Taylor expansion (i.e., Fresnel approximation) as~\cite{NFtutorial} 
\begin{equation}\label{eq:d-2nd}
d_{m,\ell}
\approx
r_\ell
-
\tilde{m} d \sin\theta_\ell
+
\frac{(\tilde{m} d)^2}{2 r_\ell}\cos^2\theta_\ell.
\end{equation}
Substituting \eqref{eq:d-2nd} into the phase expression in \eqref{eq:b-exact} yields
\begin{equation}\label{eq:phase-near}
\psi_{m,\ell}^{\rm near}
=
\tilde{m}\omega_\ell+\tilde{m}^2\phi_\ell,
\end{equation}
where  
\begin{subequations}
\label{eq:chirp-position-mapping}
\begin{align}
\omega_\ell
&=
\frac{2\pi d\sin\theta_\ell}{\lambda_{\rm c}} \in [\omega_{\min},\omega_{\max}],
\\
\phi_\ell
&=
-\frac{\pi d^2\cos^2\theta_\ell}{\lambda_{\rm c}r_\ell} \in [\phi_{\min},\phi_{\max}],
\end{align}
\end{subequations} 
where $\omega_{\min}= \frac{2\pi d \sin\theta_{\min}}{\lambda_{\rm c}}$, $\omega_{\max}=\frac{2\pi d \sin\theta_{\max}}{\lambda_{\rm c}}$, $\phi_{\min}=-\frac{\pi d^2 c_{\max}}{\lambda_{\rm c} r_{\min}}$, $\phi_{\max}=-\frac{\pi d^2 c_{\min}}{\lambda_{\rm c} r_{\max}} $, $c_{\min}
=
\min_{\theta\in[\theta_{\min},\theta_{\max}]}\cos^2\theta
$, and  $
c_{\max}
=
\max_{\theta\in[\theta_{\min},\theta_{\max}]}\cos^2\theta.$ 
Based on \eqref{eq:h-exact}-\eqref{eq:phase-near}, the received signal at the $m$-th antenna can be approximated as
\begin{equation}
\label{eq:chirp-model}
y_m
=
\sum_{\ell=1}^{L}
\alpha_\ell
e^{{\rm j}(\tilde{m}\omega_\ell + \tilde{m}^2\phi_\ell)}
+
n_m.  
\end{equation}
Eq.~\eqref{eq:chirp-model} 
reveals that the near-field array response exhibits a {\em quadratic} phase variation across the array aperture, originating from the {\em spherical} wavefront propagation in the near-field regime.
This structure is analogous to a {\em chirp signal} in the {\em time} domain, where the quadratic phase term introduces a spatially varying instantaneous frequency. 
This observation motivates the following chirp-based representation of the near-field channel~\cite{1988NFchirp,2018NFchirp}.
\begin{remark}
     From \eqref{eq:chirp-model}, the near-field channel can be viewed as a superposition of $L$ spatial chirp components in the antenna domain. Each {chirp component} is parameterized by $(\omega_\ell,\phi_\ell)$, where $\omega_\ell$ refers to the {initial spatial frequency}, $\phi_\ell$ refers to the {chirp rate}, and $L$ refers to the {model order}. Furthermore, the {position parameters} $(\theta_\ell,r_\ell)$ are uniquely mapped to the {chirp parameters} $(\omega_\ell,\phi_\ell)$ via \eqref{eq:chirp-position-mapping}. 
\end{remark}

Note that as the propagation distance $r_\ell$ increases or the array aperture $D$ decreases, the second-order phase term in \eqref{eq:phase-near} gradually becomes negligible. 
In this case, the spherical wavefront can be approximated as a planar wavefront. 
Specifically, under the Fraunhofer approximation, the first-order Taylor expansion of $d_{m,\ell}$ yields
$d_{m,\ell}
\approx
r_\ell
-
\tilde{m} d \sin\theta_\ell$. Then, the phase shift $\psi_{m,\ell}$ reduces to 
a {\em linear} phase term, given by $\psi_{m,\ell}^{\rm far}=\tilde{m}\omega_\ell$. 
Therefore, the received signal at the $m$-th antenna becomes
\begin{equation}
\label{eq:sin-model}
y_m^{\rm far} =
\sum_{\ell=1}^{L}
\alpha_\ell
e^{{\rm j}\tilde{m}\omega_\ell}
+
n_m.  
\end{equation}
This corresponds to the classical {\em sum-of-sinusoids} model widely adopted in far-field array processing~\cite{VALSE}, where each angle $\theta_\ell$ is uniquely mapped to a spatial frequency $\omega_\ell$.

\vspace{-2mm}

\subsection{Channel Parameter Estimation Problem}
\label{subsec:problem0}

In practical near-field propagation environments, the number of propagation paths is generally unknown at the BS. Moreover, the measurement noise power may vary across different operating conditions and is typically unavailable {\em a priori}, especially when only a single observation is available. Therefore, we consider a channel estimation problem in which both the number of propagation paths $L$ and the noise variance $\sigma_{\rm n}^2$ are unknown.
Based on the received signal model in \eqref{eq:y-matrix}, the objective is to estimate the geometrical parameters $\{ {\bm \alpha},{\bm \theta},{\mathbf r},L\}$
from a single observation $\mathbf y$, and then reconstruct the near-field channel based on ${\mathbf h}
    =
    \mathbf B({\bm \theta},{\mathbf r}){\bm \alpha}$.

The \ac{PDF} of $\mathbf y$ conditioned on $\{ {\bm \alpha},{\bm \theta},{\mathbf r},L \}$ is given by
\begin{align}
    p\!\left(
    \mathbf y
    \mid
    {\bm \alpha},{\bm \theta},{\mathbf r},L
    \right)
    =
    \mathcal{CN}
    \!\left(
    \mathbf y;
    \mathbf B({\bm \theta},{\mathbf r}){\bm \alpha},
    \sigma_{\rm n}^2\mathbf I
    \right).
    \label{eq:L-y}
\end{align}
Therefore, the parameter estimation can be performed based on the \ac{ML} criterion, with the log-likelihood function being
\begin{align}\label{eq:log-L-y}
    \log p\!\left(
\mathbf y
\mid 
{\bm \alpha},{\bm \theta},{\mathbf r},L
\right) \propto - 
\left\|
\mathbf y-
{\mathbf B}( {\bm \theta}, {\mathbf r})\boldsymbol\alpha
\right\|_2^2.
\end{align}
However, directly solving this ML problem is highly challenging. Even for the single-path case, the likelihood function is generally highly {\em multi-modal} \ac{w.r.t.} the angle and range parameters~\cite{yuan2024scalable}, making gradient-based optimization highly sensitive to initialization. In the multipath scenario with {\em unknown model order and noise variance}, the estimation problem becomes substantially more complicated. Exhaustive grid search over a densely discretized angle-range domain can in principle identify the global optimum, whereas its computational complexity is prohibitive for ELAA systems.

The above challenges motivate us to develop a {\em gridless} near-field channel estimation method that avoids exhaustive grid search while jointly handling the unknown path number and noise variance from a single observation.
In far-field array processing, such requirements can be effectively addressed by {\em gridless line spectral estimation} methods~\cite{VALSE}. 
However, these far-field gridless estimators cannot be directly applied to near-field parameter estimation since the near-field signal is not a sum of sinusoids but spatial chirps, as shown in \eqref{eq:chirp-model} and \eqref{eq:sin-model}.  
Interestingly, although the entire ELAA operates in the near-field regime, a sufficiently small subarray has a limited aperture such that the spherical wavefront can be locally approximated as a planar wavefront~\cite{PWFF,yuan2024scalable}. 
Motivated by this observation, we propose to appropriately partition the ELAA into multiple subarrays, enabling gridless line spectral estimation locally at each subarray.
By fusing the local estimates across subarrays, the underlying near-field channel parameters can subsequently be recovered. This array partitioning-based method is detailed in the following sections.

\vspace{-2mm}

\section{Array Partitioning and Problem Reformulation}
\label{sec:model}

In this section, we first introduce the array partitioning and the corresponding local and global coordinates.  
We then establish the \ac{CSF} model, which characterizes how a suitable array partitioning enables a local far-field sinusoidal representation within each subarray and how these local models are coupled across subarrays through the spatial chirp structure.
Based on the CSF model, we reformulate the original near-field channel estimation problem into a two-stage estimation problem.

\begin{figure}
    \centering
    \includegraphics[width=0.75\linewidth]{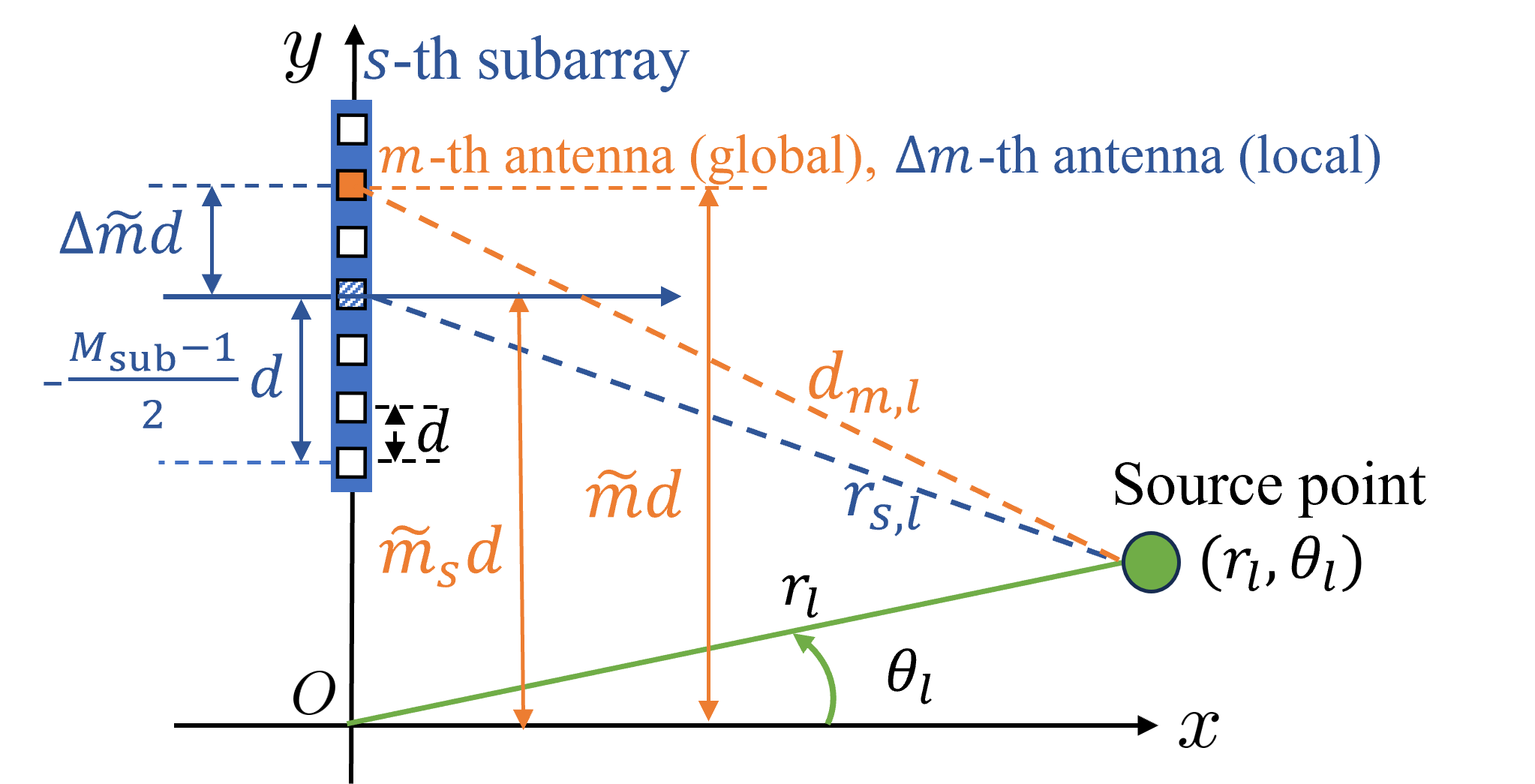}
    \vspace{-2mm}
    \caption{Global and local coordinates for the $s$-th subarray.}
    \label{fig:ULAlocal}
    \vspace{-2mm}
\end{figure}

\vspace{-2mm}

\subsection{Array Partitioning}

The ELAA is partitioned into $S$ non-overlapping subarrays, where each subarray consists of $M_{\rm sub}$ consecutive antenna elements with aperture $D_{\rm sub}=(M_{\rm sub}-1)d$.
Accordingly, the total number of antennas satisfies $M = S M_{\rm sub}$. 
Let $s \in \{0,\dots,S-1\}$ denote the subarray index.
As illustrated in Fig.~\ref{fig:ULAlocal}, the $s$-th subarray is centered at the Cartesian coordinate $[0,\tilde{m}_s d]^\mathsf{T}$, where 
\begin{align}\label{eq:tilde-m-s}
    \tilde{m}_{s} = -\frac{M-1}{2}+ s M_{\rm sub} +\frac{M_{\rm sub}-1}{2}.
\end{align}
Within each subarray, we define a local antenna index $\Delta m \in \{0, \dots, M_{\rm sub}-1\}$. 
The corresponding global antenna index of the $\Delta m$-th element in the $s$-th subarray is $m = s M_{\rm sub} + \Delta m$.
We define its local coordinate relative to the subarray center as $[0, \Delta\tilde m d]^\mathsf{T}$, where $\Delta \tilde m = \Delta m - \frac{M_{\rm sub}-1}{2}$. Its global coordinate relative to the array center is $[0, \tilde m d]^\mathsf{T}$, where
\begin{equation}\label{eq:Delta_m}
\tilde m \overset{(a)}{=} m  - (M-1)/2 \overset{(b)}{=} \tilde m_s + \Delta \tilde m,
\end{equation}
where (a) is from the definition of $\tilde m$ and (b) is from \eqref{eq:tilde-m-s}.

\vspace{-2mm}

\subsection{Chirp-Coupled Subarray Far-Field (CSF) Model}\label{subsec:CSF}

We next establish the \ac{CSF} model, which characterizes the local far-field approximation under an appropriate subarray-aperture condition and the cross-subarray coupling induced by the spatial chirp structure.

\begin{prop}
[CSF Model]\label{prop:CSF}
Suppose that the path ranges and subarray aperture satisfy 
\begin{equation}
\label{eq:subarrayFF-Rayleigh}
r_\ell \geq \frac{2D_{\rm sub}^2}{\lambda_{\rm c}}, \quad \ell=1,...L,
\end{equation} 
where $\frac{2D_{\rm sub}^2}{\lambda_{\rm c}}$ is related to the classical Rayleigh-distance criterion~\cite{selvan2017fraunhofer} associated with the subarray aperture.
The received signal at the $\Delta m$-th antenna of the $s$-th subarray can then be locally approximated as
\begin{subequations}\label{CSFmodel}
\begin{align}\label{eq:y-subarray}
y_{s,\Delta m}
=\sum_{\ell=1}^{L}
\alpha_{\ell,s} e^{{\rm j} \Delta m \omega_{s,\ell}}
+ n_{s,\Delta m},
\end{align}
where $\omega_{s,\ell}$ denotes the local spatial frequency given by
\begin{align}\label{eq:LocalFreq}
    \omega_{s,\ell}\triangleq\omega_\ell + 2\phi_\ell \tilde{m}_s,
\end{align}
\end{subequations}
$\alpha_{\ell,s}=\alpha_\ell e^{{\rm j} (\tilde{m}_s\omega_{\ell}+\tilde{m}_s^2\phi_\ell)}  e^{-{\rm j}\omega_{s,\ell} \frac{M_{\rm sub}-1}{2}} $ denotes the local path gain, and $n_{s,\Delta m}$ is the measurement noise.  
\begin{proof}
See Appendix~\ref{app:CSF}.
\end{proof}
\end{prop}

The CSF model in \eqref{CSFmodel} captures two complementary structures under the condition in \eqref{eq:subarrayFF-Rayleigh}. First, although the full-array observation exhibits {\em global near-field spatial chirps} induced by spherical wavefronts and parameterized by $\{\omega_\ell,\phi_\ell\}$, the observation within each sufficiently small subarray is represented by {\em local far-field sinusoids} induced by planar wavefronts and parameterized by $\{\omega_{s,\ell}\}$. Second, the {\em local spatial frequencies} associated with the same propagation path, i.e., $\{\omega_{s,\ell}\}_{s=0}^{S-1}$, are not independent across subarrays; rather, they are coupled through $\omega_{\ell}$ and $\phi_\ell$ in \eqref{eq:LocalFreq}.
For clarity, we refer to $\omega_\ell$ as the 
{\em global spatial frequency}. 
The cross-subarray variation of the local spatial frequencies, together with its physical interpretation, is discussed below.

\subsubsection{Cross-Subarray Spatial-Frequency Variation}\label{subsec:Freq}

\begin{figure}
\vspace{-3mm}
    \centering
    \includegraphics[width=0.75\linewidth]{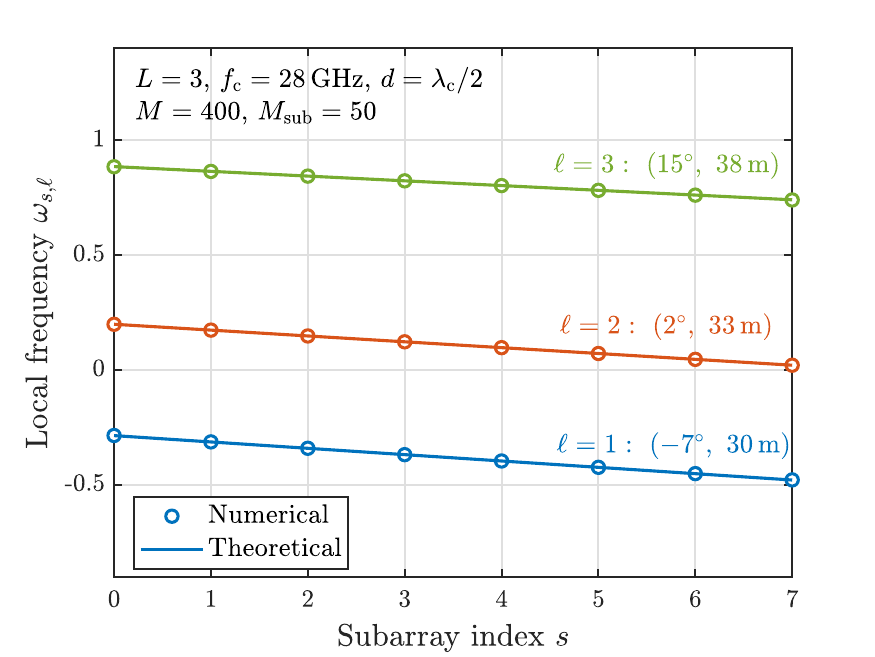}
    \vspace{-2mm}
    \caption{Spatial frequency variation across subarrays. The numerical results are obtained from the first-order derivatives of the exact phase, while the theoretical linear trajectories follow the CSF model in \eqref{CSFmodel}.} 
    \vspace{-2mm}
    \label{fig:LocalFrequencies}
\end{figure}

From \eqref{eq:chirp-model}, each physical path corresponds to a spatial chirp, parameterized by chirp parameter vector $\mathbf x_\ell=[\omega_\ell,\phi_\ell]^{\mathsf T} \in \mathcal F$, where $\mathcal F$ is the feasible set.
From \eqref{eq:chirp-position-mapping}, $\mathcal F$ is given by
\begin{align}\label{eq:feasible}
\mathcal F
&=
\big\{[\omega,\phi]^{\mathsf{T}} \!\! : 
\omega_{\min}\le\omega \le \omega_{\max}, \phi_{\min}\le\phi \le \phi_{\max}
\big\}. 
\end{align}
Under the CSF model in \eqref{CSFmodel}, for subarray $s$ with center $\tilde m_s$, this chirp component induces the local spatial frequency $\omega_{s,\ell}$, which varies across subarrays according to 
\begin{equation}\label{eq:line-vector}
\omega_{s,\ell}
=
\mathbf q_s^{\mathsf T}\mathbf x_\ell
=
\omega_\ell+2\phi_\ell\tilde m_s,
\end{equation}  
where $\bm q_s=[1,\,2\tilde m_s]^\mathsf T$.
Hence, the point set $\{(\tilde m_s,\omega_{s,\ell})\}_{s=0}^{S-1}$ lies on a straight line in the $(\tilde m,\omega)$ plane, with intercept $\omega_\ell$ and slope $2\phi_\ell$, as shown in Fig.~\ref{fig:LocalFrequencies}. We refer to this plane as the {\em subarray-center--spatial-frequency plane} and this straight line as the {\em chirp trajectory} associated with path $\ell$.  

This linear variation follows from the quadratic phase profile of the global near-field spatial chirp: The local spatial frequency corresponds to the first-order phase slope evaluated at each subarray center, as illustrated in Fig.~\ref{fig:SubarrayAoA}. Therefore, the spherical-wave curvature is preserved through the linear variation of local spatial frequencies across subarrays.

\subsubsection{Local-Angle Interpretation}

\begin{figure}
\vspace{-2mm}
    \centering
    \includegraphics[width=0.7\linewidth]{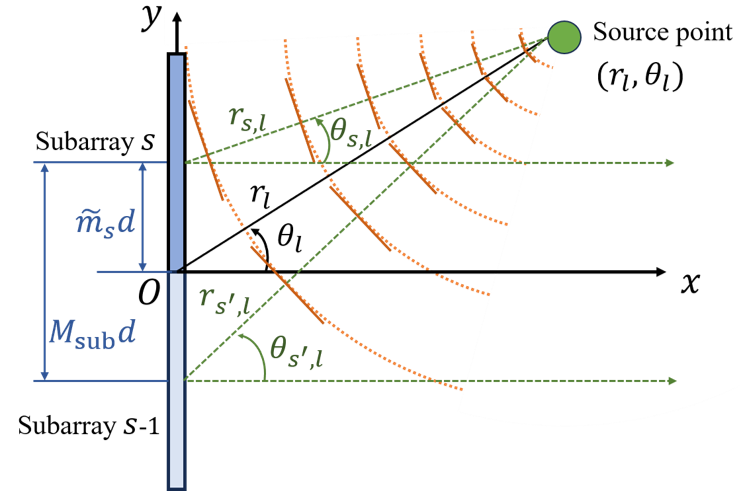}
    \vspace{-1mm}
    \caption{Illustration of the CSF model. The spherical-wave curvature gives rise to locally planar wavefronts at different subarrays, resulting in spatially varying local angles $\theta_{s,\ell}$ relative to the global angle $\theta_\ell$.} \label{fig:SubarrayAoA}
    \vspace{-2mm}
\end{figure}

Note that, for path $\ell$, the {\em global spatial frequency} $\omega_\ell$ corresponds to the physical angle $\theta_\ell$ observed at the full-array center (referred to as {\em global angle}) via the mapping in  \eqref{eq:chirp-position-mapping}.
Extending this mapping to subarrays, the {\em local spatial frequency} $\omega_{s,\ell}$ naturally defines a {\em local angle} observed at the $s$-th subarray center, denoted by $\theta_{s,\ell}$, i.e., 
\begin{align}\label{eq:LocalFreq-angle}
    \omega_{s,\ell} = \frac{2\pi d \sin\theta_{s,\ell}}{\lambda_{\rm c}} .
\end{align} 
The variation of local spatial frequency across subarrays in \eqref{eq:LocalFreq} thus implies the variation of local angles. 
As visualized in Fig.~\ref{fig:SubarrayAoA}, each subarray observes a locally planar wave whose apparent angle is slightly different from the global angle at the array center.
The following proposition formalizes the variation of local angles across subarrays.

\begin{prop}[Local-Angle Variation]
\label{prop:local_angle} 
Under the CSF model, for the $\ell$-th path, the local angle $\theta_{s,\ell}$ at the $s$-th subarray is related to the global angle $\theta_\ell$ by
\begin{align}\label{eq:LocalAngle}
\sin\theta_{s,\ell}
=
\sin\theta_{\ell}
-
\frac{d \tilde{m}_s}{r_\ell}\cos^2\theta_\ell.
\end{align}

\vspace{-3mm}

\begin{proof}
Eq.~\eqref{eq:LocalAngle} can be derived in two equivalent ways.
First, it follows directly from the mapping between the spatial frequency and physical angle in \eqref{eq:chirp-position-mapping} and \eqref{eq:LocalFreq-angle}, together with the relationship between the local and global spatial frequencies in \eqref{eq:LocalFreq}.  
Second, it follows from a geometric viewpoint by performing a {\em first-order} expansion of the source direction observed at the subarray center. 
The detailed derivation is provided in Appendix~\ref{app:subarrayAoA}.
\end{proof}
\end{prop}

Eq.~\eqref{eq:LocalAngle} shows that the sine of the local angle, i.e., $\sin\theta_{s,\ell}$, varies linearly with the subarray center $\tilde m_s$. 
This variation quantifies the gradual rotation of the locally planar wavefronts induced by the global spherical wave. In the far-field limit, where $r_\ell$ is sufficiently large, the second term in \eqref{eq:LocalAngle} vanishes, and all antennas observe the same angle, i.e., $\theta_{s,\ell}\approx\theta_\ell$.
This highlights a fundamental distinction between near-field and far-field propagation. In the near-field, the local angle exhibits a {\em first-order} spatial variation \ac{w.r.t.} the subarray center, whereas in the far-field, this variation vanishes and the angle remains approximately constant, corresponding to a {\em zeroth-order} spatial approximation.

This cross-subarray angle variation also reflects near-field angular dispersion. 
Specifically, due to the spherical wavefront, a single path $\ell$, corresponding to one physical source location $[\theta_\ell,r_\ell]^{\mathsf T}$, is observed at different local angles ${\theta_{s,\ell}}$ across the array aperture, thereby spanning a finite angular range. Such angular dispersion disappears under the far-field planar-wavefront approximation.

\vspace{-2mm}

\subsection{Two-Stage Estimation Under CSF Model}\label{subsec:reformulation}

Based on the CSF model, the original near-field angular-range estimation problem can be reformulated as the following two-stage procedure.

\begin{itemize}
\item \textbf{Gridless Subarray-Level Inference:} 
Estimate the local spatial frequencies $\{\omega_{s,\ell}\}$ independently for each subarray under the far-field sinusoidal model in \eqref{eq:y-subarray}. 
By partitioning the array, the original single near-field observation is transformed into multiple subarray observations, each corresponding to a standard line spectral estimation problem. 
A gridless estimator such as VALSE can be employed to infer the local spatial frequencies directly from a single subarray snapshot without requiring prior knowledge of the path number or measurement noise variance. 
\item \textbf{Cross-Subarray Fusion:}
Recover the underlying chirp trajectories from subarray-level estimates. By exploiting the linear variation of local spatial frequencies across subarrays in \eqref{eq:line-vector} under the CSF model, the slope and intercept of each recovered linear trajectory yield the chirp parameters $(\omega_\ell,\phi_\ell)$, which are subsequently mapped to the position parameters $(\theta_\ell,r_\ell)$ through \eqref{eq:chirp-position-mapping}. 

\end{itemize}

This two-stage reformulation leads to a {\em structured local-inference and global-fusion} problem.
However, the local estimates are obtained independently from each subarray and subject to estimation uncertainty, thereby forming an {\em unordered} set of imperfect local estimates without explicit {\em path association} information. 
Consequently, reliable cross-subarray association is required to recover the underlying chirp trajectories. 
To this end, the following sections develop practical algorithms to address these challenges, thereby enabling accurate estimation of both the path number and near-field angular-range parameters.

\section{CHARM Algorithm Under CSF Model}\label{sec:CHARM}

Building on the two-stage estimation framework introduced in Sec.~\ref{subsec:reformulation}, we propose the \ac{CHARM} algorithm in this section.

\vspace{-2mm}

\subsection{Gridless Subarray-Level Inference}

As the first stage of CHARM, gridless local estimation is performed independently for each subarray under the CSF model in \eqref{eq:y-subarray}.
By stacking the received signals of all antennas within subarray $s$ into a vector $\mathbf y_s
=
[y_{s,0},y_{s,1},\ldots,y_{s,M_{\rm sub}-1}]^{\mathsf T}$, the subarray observation can be expressed as the standard line spectral model by
\begin{equation}
\mathbf y_s
=
\sum_{\ell=1}^{L_s}
\alpha_{s,\ell}\mathbf a(\omega_{s,\ell})
+
\mathbf n_s,
\end{equation} 
where $\mathbf a(\omega)
=
[1,e^{{\rm j}\omega},\dots,e^{{\rm j}(M_{\rm sub}-1)\omega}]^{\rm T}$, $\omega_{s,\ell}$ is the spatial frequency, $\alpha_{s,\ell}$ is the path gain, $L_s=L$ is the model order, and 
$\mathbf n_s =
[n_{s,0},n_{s,1},\ldots,n_{s,M_{\rm sub}-1}]^{\mathsf T}$.

We apply VALSE~\cite{VALSE} independently to each subarray for gridless line spectral estimation and automatic model-order determination, with $L_{\rm valse}$ denoting the prescribed maximum number of candidate spectral components. For the components inferred to be present, their path gains are modeled by a zero-mean complex Gaussian prior with unknown variance $\tau_s$.
VALSE yields the estimated model order $\hat L_s\leq L_{\rm valse}$. Moreover, for each detected component $i=1,\ldots,\hat L_s$, VALSE provides the local spatial-frequency and path-gain estimates, denoted by $\hat\omega_{s,i}$ and $\hat\alpha_{s,i}$, respectively.
It also provides the corresponding uncertainty measures
$\sigma_{\omega,s,i}^2$ and $\sigma_{\alpha,s,i}^2$, together with the
estimated path-gain prior variance $\hat\tau_s$ and noise power
$\hat\sigma_{{\rm n},s}^2$. 
We define the local observation set at subarray $s$ as
\begin{equation}
    \mathcal Z_s
    =
    \left\{
    \big(\hat\omega_{s,i},\hat\alpha_{s,i}\big)
    \right\}_{i=1}^{\hat L_s},
\end{equation}
and collect the observations from all subarrays as $\mathcal Z = \left\{  \mathcal Z_s \right\}_{s=0}^{S-1}$.
Thus, the original received signal $\mathbf y$ is transformed into the local observation set $\mathcal Z$ and its associated statistical information, which serves as the input to the subsequent cross-subarray fusion stage.

\vspace{-2mm}

\subsection{Cross-Subarray Fusion}
\label{subsec:cross-subarray-fusion}
 
\begin{figure}
\vspace{-3mm}
    \centering
\includegraphics[width=0.75\linewidth]{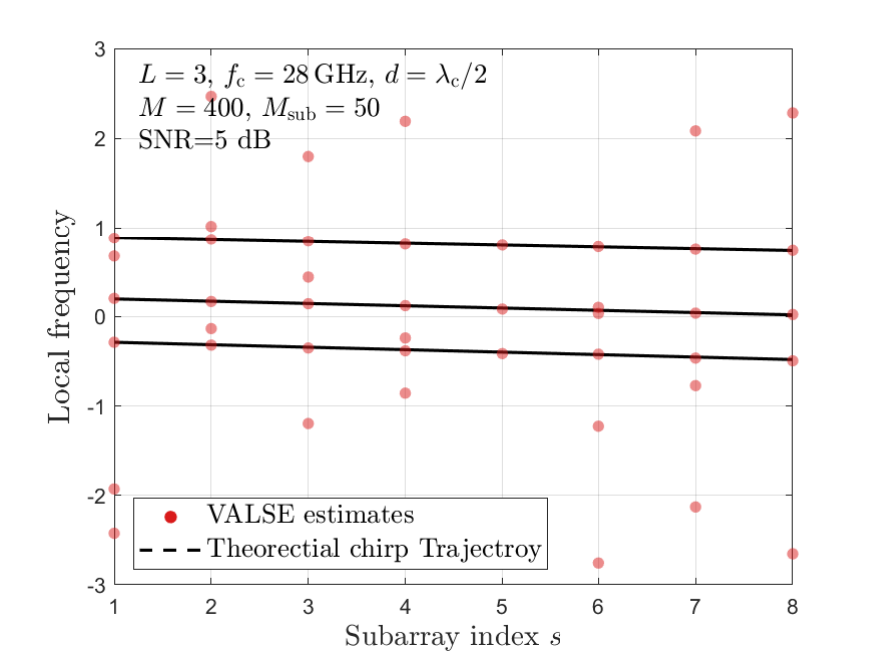}
\vspace{-2mm}
    \caption{Local frequency estimates from subarray-level VALSE.}
    \vspace{-2mm}
    \label{fig:valse}
\end{figure}

The objective of the second stage is to fuse the local observations in $\mathcal Z$ to recover the underlying chirp trajectories.
As discussed in Sec.~\ref{subsec:Freq}, the local spatial frequencies generated by the same physical propagation path vary linearly across subarrays, forming a path-specific chirp trajectory in \eqref{eq:line-vector}.
Therefore, the local frequency observations obtained from different subarrays can be interpreted as {\em noisy samples} drawn from multiple chirp trajectories, as shown in Fig.~\ref{fig:valse}. 
Hence, the cross-subarray fusion can be formulated as a {\em multi-trajectory fitting} problem.

However, this fitting problem is nontrivial. 
Since VALSE is performed independently for each subarray,  the local component indices, e.g., $i$ in $\hat\omega_{s,i}$ and $\hat\omega_{s',i}$ for different subarrays $s,s'\in\{0,\dots,S-1\}$, do not necessarily correspond to the same physical path, resulting in unknown cross-subarray path {\em associations}. 
Moreover, measurement noise and inter-path correlation may introduce {\em missed detections} and {\em spurious components} in the local observations, such that the estimated local model order $\hat L_s$ may vary across subarrays and does not necessarily equal the number of underlying physical paths $L$.
Therefore, the trajectory fitting problem needs to jointly resolve the {\em unknown observation-to-trajectory associations} and account for {\em unreliable local observations} with inconsistent model orders.
To address these issues, we formulate cross-subarray fusion as a probabilistic multi-trajectory fitting problem and solve it using an \ac{EM} framework.

\subsubsection{Probabilistic Problem Formulation}
We assume $L_{\rm em}$ candidate chirp trajectories, where the $\ell$-th trajectory is parameterized by the chirp parameter vector $\mathbf x_\ell$, $\ell=1,\ldots,L_{\rm em}$. 
The corresponding chirp parameter set is denoted by $\mathcal X = \left\{ \mathbf x_\ell = [\omega_\ell,\phi_\ell]^{\mathsf T} \right\}_{\ell=1}^{L_{\rm em}}$.
According to \eqref{eq:line-vector}, the local spatial frequency generated by the $\ell$-th trajectory at subarray $s$ is $\mathbf q_s^{\mathsf T}\mathbf x_\ell$.  
Given the local observation set $\mathcal Z$, our objective is to infer the underlying chirp parameter set $\mathcal X$ based on \ac{ML}, i.e., 
\begin{align}\label{eq:fusion-ML}
\hat{\mathcal X}
=
\arg\max_{\mathcal X}\quad
&\log p(\mathcal Z\mid\mathcal X) 
\\
\nonumber\text{s.t.}\quad
&\mathbf x_\ell\in\mathcal F,
\quad \ell=1,\ldots,L_{\rm em}.
\end{align}
where $\mathcal F$ is defined in \eqref{eq:feasible}.

Since the correspondence between the local observations and the candidate trajectories is unknown, the likelihood $p(\mathcal Z\mid\mathcal X)$ implicitly involves marginalization over all possible observation-to-trajectory associations. 
To explicitly represent these hidden associations, 
for each local observation indexed by $(s,i)$, we introduce a latent association variable $c_{s,i}\in\{0,1,\dots,L_{\rm em}\}$, 
where $c_{s,i}=\ell\in\{1,\ldots,L_{\rm em}\}$ indicates that the $(s,i)$-th local observation belongs to the $\ell$-th chirp trajectory, while $c_{s,i}=0$ corresponds to a spurious component.
Without additional prior information, a uniform association prior is adopted, i.e., 
\begin{equation}\label{eq:prior-asso}
p(c_{s,i}=\ell)
=
\frac{1}{L_{\rm em}+1},
\qquad
\ell=0,\ldots,L_{\rm em}.
\end{equation}
Let $\mathcal C = \left\{ \{c_{s,i}\}_{i=1}^{\hat L_s} \right\}_{s=0}^{S-1}$. We can rewrite $p(\mathcal Z\mid\mathcal X)$ as
\begin{align}\label{eq:marginal-C} 
p(\mathcal Z\mid\mathcal X) &= \sum_{\mathcal C} p(\mathcal Z,\mathcal C\mid\mathcal X) = \sum_{\mathcal C} p(\mathcal C)\, p(\mathcal Z\mid\mathcal C,\mathcal X)
\nonumber\\& \overset{(a)}{=} \prod_{s=0}^{S-1} \prod_{i=1}^{\hat L_s} \sum_{\ell=0}^{L_{\rm em}} p(c_{s,i}=\ell) p(\hat \omega_{s,i},\hat \alpha_{s,i}\mid c_{s,i}=\ell,\mathcal X)
\nonumber\\& \overset{(b)}{=} \prod_{s=0}^{S-1} \prod_{i=1}^{\hat L_s} \sum_{\ell=0}^{L_{\rm em}}  p(c_{s,i}=\ell) p(\hat \omega_{s,i} \mid c_{s,i}=\ell,\mathcal X) \times\nonumber\\& \qquad\qquad  p(\hat \alpha_{s,i} \mid c_{s,i}=\ell,\mathcal X),
\end{align}
where (a) and (b) follow the independence across and within the local observations, respectively. 

We next specify the likelihood of each local observation. 
Conditioned on a valid association $c_{s,i}=\ell$, $\ell\geq1$, the local spatial-frequency estimate is modeled as a noisy observation of the corresponding chirp trajectory in \eqref{eq:line-vector}. 
Approximating the frequency-estimation error as Gaussian with variance given by the VALSE uncertainty measure $\sigma_{\omega,s,i}^2$, we model $\hat\omega_{s,i}$ as
\begin{equation}\label{eq:freq-likelihood} p(\hat\omega_{s,i}\mid c_{s,i}=\ell,\mathcal X) = \mathcal N\!\left( \hat\omega_{s,i}; \mathbf q_s^{\mathsf T}\mathbf x_\ell, \sigma_{\omega,s,i}^2 \right), \ell\geq1.
\end{equation} 
For a spurious component, no chirp-induced line structure is imposed, and the local spatial frequency is modeled as uniformly distributed over the spatial-frequency domain, i.e., 
\begin{equation}\label{eq:freq-likelihood-spurious} p(\hat\omega_{s,i}\mid c_{s,i}=0,\mathcal X) = \mathcal U(\hat\omega_{s,i};-\pi,\pi). 
\end{equation}
The local path-gain estimate provides complementary information for distinguishing valid and spurious components.
For a valid component $\ell\ge 1$, conditioned on $c_{s,i}=\ell$, the local path-gain estimate is modeled as
$\hat\alpha_{s,i} = \alpha_{s,\ell} + \varepsilon_{\alpha,s,i}$, 
where  $\alpha_{s,\ell}\sim\mathcal{CN}(0,\hat\tau_s)$ follows the path-gain prior adopted in VALSE~\cite{VALSE}, and $\varepsilon_{\alpha,s,i} \sim \mathcal{CN}(0,\sigma_{\alpha,s,i}^2)$ accounts for the gain-estimation error. Therefore,
\begin{equation}\label{eq:gain-likelihood} p(\hat\alpha_{s,i}\mid c_{s,i}=\ell,\mathcal X) = \mathcal{CN}\!\left( \hat\alpha_{s,i}; 0, \hat\tau_s+\sigma_{\alpha,s,i}^2 \right), \ell\geq1. \end{equation} 
For $c_{s,i}=0$, no valid path gain is present,
and the local gain estimate is modeled solely by the estimation error. Therefore, \begin{equation}\label{eq:gain-likelihood-spurious} p(\hat\alpha_{s,i}\mid c_{s,i}=0,\mathcal X) = \mathcal{CN}\!\left( \hat\alpha_{s,i}; 0, \sigma_{\alpha,s,i}^2 \right). \end{equation}
Together, the spatial-frequency likelihood evaluates the consistency of each local observation with a candidate chirp trajectory while accounting for its frequency-estimation uncertainty, whereas the path-gain likelihood provides complementary information for suppressing spurious local components.

Substituting the above likelihoods into \eqref{eq:marginal-C} fully specifies the ML objective in \eqref{eq:fusion-ML}. 
However, the resulting log-sum structure in $\log p(\mathcal Z\mid\mathcal X)$, induced by the unknown latent associations, makes direct optimization of \eqref{eq:fusion-ML} nontrivial, which motivates the EM-based solution developed next.

\subsubsection{EM-Based Solution}

To handle the latent association variables, we employ the \ac{EM}
algorithm~\cite{dempster1977EM}.
In the EM formulation, $\mathcal Z$ constitutes the observed data,
$\mathcal C$ represents the latent data, and
$(\mathcal Z,\mathcal C)$ is regarded as the complete data.
The EM algorithm then alternates between inferring the latent
associations in the E-step and updating the chirp parameters in the M-step.

Denote the chirp parameter estimate at the $t$-th iteration by $ \mathcal X^{(t)}=\left\{\mathbf x_\ell^{(t)}=[\omega_\ell^{(t)},\phi_\ell^{(t)}]^\mathsf{T}\right\}_{\ell=1}^{L_{\rm em}}$.
At the $(t+1)$-th iteration, the {\em E-step} computes the posterior distributions of the latent association variables under the current chirp parameter estimate, i.e., $p( \mathcal C \mid \mathcal Z, \mathcal X^{(t)})$. Then, the {\em M-step} updates the chirp parameter estimate by maximizing the expected complete-data log-likelihood under the inferred soft associations, i.e., $\mathbb E_{ \mathcal C \mid \mathcal Z, \mathcal X^{(t)} } \!\left[ \log p( \mathcal Z, \mathcal C \mid \mathcal X) \right]$. 
In the following, we detail the E-step and M-step at the $(t+1)$-th iteration in Sec.~\ref{subsec:E-step} and Sec.~\ref{subsec:M-step}, respectively.  
Moreover, a reliable initialization strategy for $\mathcal X^{(0)}$ is introduced in Sec.~\ref{subsec:RANSAC}.

\paragraph{Soft Association in E-Step}\label{subsec:E-step}

Given the current chirp parameter estimate $\mathcal X^{(t)}$, the E-step computes the posterior probability that the $(s,i)$-th local observation is associated with the $\ell$-th candidate chirp trajectory.
We denote the posterior association probability in the $(t+1)$-th iteration by
\begin{equation}\label{eq:responsibility0}
\!\varrho _{s,i,\ell}^{(t+1)}
\triangleq
p(c_{s,i}\!=\ell
\mid
\hat\omega_{s,i},\hat\alpha_{s,i},\mathcal X^{(t)}), \ell=0,1,\ldots,L_{\rm em}.
\end{equation}
By Bayes' rule, \eqref{eq:responsibility0} can be further expressed as
\begin{align}\label{eq:responsibility}
  \!\varrho _{s,i,\ell}^{(t+1)}
\!=\!\frac{
p(c_{s,i}\!=\ell)\,
p(\hat\omega_{s,i},\hat\alpha_{s,i}\mid c_{s,i}\!=\ell,\mathcal X^{(t)})
}{
\sum_{\ell'=0}^{L_{\rm em}}
p(c_{s,i}\!=\ell')\,
p(\hat\omega_{s,i},\hat\alpha_{s,i} \mid c_{s,i}\!=\ell',\mathcal X^{(t)})
},\!\!
\end{align}
where the prior term $p(c_{s,i}=\ell)$ is defined in \eqref{eq:prior-asso}, and the likelihood term is evaluated using \eqref{eq:marginal-C}-\eqref{eq:gain-likelihood-spurious}.
Therefore, $\varrho_{s,i,\ell}^{(t+1)}$ quantifies the soft association confidence between the $(s,i)$-th local observation and the $\ell$-th candidate trajectory.

\paragraph{Chirp Parameter Update (M-Step)}\label{subsec:M-step}

Given the posterior association probabilities obtained in the E-step, the M-step updates the chirp parameter set by maximizing the expected complete-data log-likelihood.
Since the association prior and the path-gain likelihood are independent of $\mathcal{X}$,
only the spatial-frequency likelihood contributes to the update of the chirp parameters.
Therefore, maximizing the expected complete-data log-likelihood is equivalent to solving the following constrained \ac{WLS} problem for each candidate trajectory ($\ell\ge 1$), given by
\begin{align}\label{eq:wls-objective}
\quad \mathbf x_\ell^{(t+1)} 
= \arg\min_{\mathbf x_\ell\in\mathcal F} \sum_{s=0}^{S-1}
\sum_{i=1}^{\hat L_s}
\frac{\varrho_{s,i,\ell}^{(t+1)}}{\sigma_{\omega,s,i}^2}
\left(
\hat\omega_{s,i}
-
\mathbf q_s^{\mathsf T}\mathbf x_\ell
\right)^2,
\end{align} 
where $\ell\in\{1,\dots,L_{\rm em}\}$, the weight  $\varrho_{s,i,\ell}^{(t+1)}/\sigma_{\omega,s,i}^2$ jointly accounts for the soft association confidence and the reliability of the local frequency estimate, and $\mathcal F$ is given in \eqref{eq:feasible}.
Thus, the update of each candidate trajectory in the M-step reduces to a two-dimensional box-constrained convex quadratic program, which can be efficiently solved using standard convex optimization tools.

\paragraph{RANSAC for EM Initialization}
\label{subsec:RANSAC}

The performance of the EM algorithm depends on the initialization of the chirp parameter set $\mathcal X^{(0)}$. Poor initialization may lead to incorrect associations in the E-step and cause the subsequent iterations to converge to undesirable local optima.
To obtain reliable initial chirp trajectories, we exploit the linear variation of the local spatial frequencies across subarrays in \eqref{eq:line-vector}.
As discussed above, the local frequency estimates in $\mathcal Z$ form a noisy point set around multiple underlying lines in the $(\tilde m,\omega)$ plane.
This motivates the use of \ac{RANSAC}~\cite{fischler1981ransac}, a robust model-fitting approach that repeatedly generates candidate models from randomly sampled observations and identifies well-supported models in the presence of outliers.
In our problem, each chirp trajectory corresponds to a straight-line model in the $(\tilde m,\omega)$ plane.
Therefore, the line parameters obtained from RANSAC can be used to initialize the chirp parameter set $\mathcal X^{(0)}$, with $L_{\rm em}=|\mathcal X^{(0)}|$.

\vspace{-2mm}
 
\subsection{Outputs of CHARM Algorithm}
Let $T_{\rm em}$ denote the maximum number of EM iterations.
The EM output consists of the estimated chirp parameter set $\mathcal X^{(T_{\rm em})}$, and the posterior association probability set is $\{\varrho_{s,i,\ell}^{(T_{\rm em})}\}$. 
Since the $L_{\rm em}$ candidate trajectories may contain redundant or spurious components, a trajectory pruning and merging procedure is performed before channel parameter recovery.

First, the posterior association probability of the $\ell$-th candidate trajectory is aggregated across all subarrays as
\begin{equation}
\varrho_\ell
=
\sum_{s=0}^{S-1}
\sum_{i=1}^{\hat L_s}
\varrho_{s,i,\ell}^{(T_{\rm em})},
\qquad
\ell=1,\dots,L_{\rm em},
\end{equation}
which is further normalized as $\bar\varrho_\ell
=
\frac{\varrho_\ell}
{\sum_{\ell'=1}^{L_{\rm em}}\varrho_{\ell'}}$.
Second, candidate trajectories with sufficiently close chirp parameters are merged.
For example, ${\mathbf x}_{\ell_1}^{(T_{\rm em})}, {\mathbf x}_{\ell_2}^{(T_{\rm em})}\in \mathcal{X}^{(T_{\rm em})}$, $\ell_1\neq\ell_2$, are regarded as redundant if
\begin{equation}\label{eq:similar}
|\omega_{\ell_1}^{(T_{\rm em})}-\omega_{\ell_2}^{(T_{\rm em})}| < \epsilon_\omega \text{ and }
|\phi_{\ell_1}^{(T_{\rm em})}-\phi_{\ell_2}^{(T_{\rm em})}| < \epsilon_\phi,
\end{equation}
where $\epsilon_\omega$ and $\epsilon_\phi$ are predefined merging thresholds.
In this case, only the trajectory with the larger association probability is retained. Third, based on the minimum association probability threshold $\epsilon_{\rm asso}$, 
trajectories satisfying $\bar\varrho_\ell < \epsilon_{\rm asso}$
are regarded as spurious and are thus removed. 
The final chirp parameter estimates are denoted by 
$\hat{\bm\omega}=[\hat\omega_1,\dots,\hat\omega_{\hat L}]^\mathsf T$ and $
\hat{\bm\phi}=[\hat\phi_1,\dots,\hat\phi_{\hat L}]^\mathsf T$, 
where $\hat L$ is the estimated path number. 

Based on the estimated chirp parameters, the mapping in \eqref{eq:chirp-position-mapping} yields the angle and range estimates, denoted by $\hat{\bm\theta}
=[\hat\theta_1,\dots,\hat\theta_{\hat L}]^\mathsf T $ and $
\hat{\mathbf r}= [\hat r_1,\dots,\hat r_{\hat L}]^\mathsf T$.
Given $\hat{\bm\theta}$, $\hat{\mathbf r}$, and $\hat L$, the complex path gains can be estimated by the \ac{ML} criterion as
\begin{align}\label{eq:alpha}
\hat{\boldsymbol\alpha} 
\overset{(a)}{=}\arg\min_{\boldsymbol\alpha}\left\|
\mathbf y
-
{\mathbf B}(\hat{\bm \theta},\hat{\mathbf r})\boldsymbol\alpha
\right\|_2^2
= \left(
\hat{\mathbf B}^{\mathsf H}\hat{\mathbf B}
\right)^{-1}
\hat{\mathbf B}^{\mathsf H}\mathbf y,
\end{align}
where (a) is from \eqref{eq:L-y}-\eqref{eq:log-L-y}, $\hat{\mathbf B}\triangleq{\mathbf B}(\hat{\bm \theta},\hat{\mathbf r})$, and ${\mathbf B}(\hat{\bm \theta},\hat{\mathbf r})$ is defined in \eqref{eq:y-matrix}.  
Accordingly, the near-field channel is reconstructed via \eqref{eq:h-exact}-\eqref{eq:y-matrix}, denoted by $\hat{\mathbf h}
=
\hat{\mathbf B}\hat{\boldsymbol\alpha}$.
The final outputs of CHARM are $\left\{
\hat L,
\hat{\boldsymbol\alpha},
\hat{\bm\theta},\hat{\mathbf r},
\hat{\mathbf h}
\right\}$. We summarize the overall procedure of CHARM in Algorithm~\ref{alg:CHARM}.

\begin{algorithm}[t]
\caption{CHARM Algorithm}
\label{alg:CHARM}
\small
\begin{algorithmic}[1]

\REQUIRE $\mathbf y$, $S$, $M_{\rm sub}$, $M$, $L_{\rm valse}$,  $\mathcal F$,  $T_{\rm em}$, $\epsilon_{\omega}$, $\epsilon_{\phi}$, and $\epsilon_{\rm asso}$

\ENSURE $\hat L$, $\hat{\boldsymbol\alpha}$, $\hat{\boldsymbol\theta}$, $\hat{\mathbf r}$, and $\hat{\mathbf h}$

\STATE Partition the ELAA into $S$ subarrays and obtain $\{\mathbf y_s\}_{s=0}^{S-1}$
 
\STATE \textbf{Stage 1: Gridless Subarray-Level Estimation}
\FOR{each subarray $s$}
    \STATE Apply \ac{VALSE} to $\mathbf y_s$
    \STATE  Obtain local estimates and their uncertainties 
\ENDFOR 
\STATE 
Collect all local observations as $\mathcal Z$

\STATE \textbf{Stage 2: EM-Based Cross-Subarray Fusion}
\STATE RANSAC-based initialization to obtain $\mathcal{X}^{(0)}$ 
\FOR{$t=0,\ldots,T_{\rm em}-1$}
    \STATE E-step: compute  posterior association probabilities $\{\varrho_{s,i,\ell}^{(t+1)}\}_{s,i,\ell}$ under $\mathcal{X}^{(t)}$ via \eqref{eq:responsibility} 
    \STATE M-step: update chirp parameter set $\mathcal{X}^{(t+1)}$ via \eqref{eq:wls-objective} 
\ENDFOR

\vspace{0.2cm}

\STATE Merge redundant trajectories in $\mathcal X^{(T_{\rm em})}$ and prune weakly associated trajectories based on $\{\varrho_{s,i,\ell}^{(T_{\rm em})}\}_{s,i,\ell}$

\STATE Obtain chirp parameters $\{(\hat\omega_\ell,\hat\phi_\ell)\}_{\ell=1}^{\hat L}$

\STATE Recover $(\hat{\boldsymbol\theta},\hat{\mathbf r})$ from
$(\hat{\boldsymbol\omega},\hat{\boldsymbol\phi})$ using \eqref{eq:chirp-position-mapping}

\STATE Estimate $\hat{\boldsymbol\alpha}$ by \eqref{eq:alpha} and reconstruct
$\hat{\mathbf h}$

\RETURN $\hat L$, $\hat{\boldsymbol\alpha}$, $\hat{\boldsymbol\theta}$, $\hat{\mathbf r}$, and $\hat{\mathbf h}$

\end{algorithmic}
\end{algorithm}

\vspace{-2mm}

\subsection{Computational Complexity}
\label{subsec:CHARM-complexity}
 
The computational complexity of CHARM mainly consists of the subarray-level VALSE processing and the EM-based cross-subarray fusion. 
For the subarray-level processing, VALSE is independently applied to each of the $S$ subarrays.
The main complexity per VALSE iteration is $\mathcal O(M_{\rm sub}L_{\rm valse}^{3})$~\cite{VALSE}, where $L_{\rm valse}$ is the prescribed maximum number of paths considered by VALSE at each subarray and $M_{\rm sub}$ is the subarray antenna number.
With $T_{\rm valse}$ iterations, the overall complexity of subarray-level processing is $\mathcal O(T_{\rm valse}SM_{\rm sub}L_{\rm valse}^{3})=\mathcal O(MT_{\rm valse}L_{\rm valse}^{3})$. 
The EM module has $T_{\rm em}$ iterations. In each iteration, the E-step evaluates the association likelihoods between all local observations and $L_{\rm em}$ candidate trajectories, while the M-step updates trajectory parameters. 
Since the number of local observations is bounded by $
\sum_{s=0}^{S-1}\hat L_s\le S L_{\rm valse}$, the resulting complexity is $\mathcal O(S L_{\rm valse}T_{\rm em} L_{\rm em} )$.

Since $L_{\rm valse}$ and $L_{\rm em}$ are both on the order of the path number $L$, the overall complexity of CHARM can be approximated as $\mathcal O \Big(
T_{\rm valse}ML^3
+ T_{\rm em}SL^2
\Big)$.
Moreover, since $S=M/M_{\rm sub}$, the CHARM complexity grows linearly with the array size $M$ when the subarray size is fixed. In addition, the subarray-level VALSE processing can be performed independently across subarrays and is therefore naturally parallelizable, making CHARM well-suited for ELAA near-field channel estimation.

\section{Enhanced CHARM Algorithm}\label{sec:E-CHARM}

Building upon the CSF model in \eqref{CSFmodel}, CHARM transforms the near-field channel estimation problem into a chirp-parameter estimation problem and recovers the path number and position parameters.
However, the CSF model is an approximation of the near-field model in \eqref{eq:y-matrix}.  The resulting modeling mismatch may limit the ultimate estimation accuracy. 
Therefore, we develop E-CHARM to further refine the CHARM estimates under the near-field model in \eqref{eq:y-matrix}.

\vspace{-2mm}

\subsection{Problem Formulation Under Near-Field Model} 

From \eqref{eq:log-L-y}, given the received signal in \eqref{eq:y-matrix}, the \ac{ML} estimation problem of the complex channel gains, the position parameters, and the path number can be formulated as
\begin{align}\label{eq:ML}
      (\hat{\boldsymbol\alpha}, \hat{\bm\theta}, \hat{\mathbf r},\hat L)=  & \arg \min_{ {\boldsymbol\alpha},  {\bm\theta},  {\mathbf r}, L} \,\left\| \mathbf y - {\mathbf B}({\bm \theta}, {\mathbf r})\boldsymbol\alpha \right\|_2^2
      \\ & \text{s.t.} ~ {\boldsymbol\alpha} \in \mathbb{C}^{L},  {\bm\theta} \in \mathbb{R}^{L},  {\mathbf r} \in \mathbb{R}^{L},
      \nonumber\\& ~~\quad \theta_{\min}\le\theta_\ell \le \theta_{\max}, ~
      r_{\min}\le r_\ell \le r_{\max}.\nonumber 
  \end{align} 
Given $\bm \theta$, $\mathbf r$, and $L$, the optimization problem in \eqref{eq:ML} is linear \ac{w.r.t.} $\bm \alpha$ with the solution given in \eqref{eq:alpha}. Substituting \eqref{eq:alpha} into \eqref{eq:ML}, the objective function in \eqref{eq:ML} can be reformulated as 
\begin{align}\label{eq:cost}
\mathcal{J}(\bm\theta,\mathbf r)=
\left\|
\left(
\mathbf I-
\mathbf P(\bm\theta,\mathbf r)
\right)
\mathbf y
\right\|_2^2,
\end{align} 
where $\mathbf P(\bm\theta,\mathbf r)
=
\mathbf B
\left(
\mathbf B^{\mathsf H}
\mathbf B
\right)^{-1}
\mathbf B^{\mathsf H}
$, 
and the model order $L$ is implicitly represented by the dimension of $\bm\theta$.

\vspace{-2mm}
 
\subsection{Algorithm Design}

Since the above optimization is nonconvex, E-CHARM is initialized with the CHARM estimate and adopts an alternating refinement strategy, where the angle vector $\bm\theta$ and range vector $\mathbf r$ are updated sequentially via gradient descent. 
Specifically, let
$\bm\theta^{(0)}$
and
$\mathbf r^{(0)}$
denote the angle-range estimates obtained from CHARM. Moreover, $\nabla_{\bm\theta}
\mathcal J$ and $\nabla_{\mathbf r}
\mathcal J$ denote the gradients of $\mathcal J$ \ac{w.r.t.} $\bm\theta$ and $\mathbf r$, respectively.
At the $(t+1)$-th iteration, the angle parameters are updated as
\begin{equation}
\bm\theta^{(t+1)}
=
\bm\theta^{(t)}
-
\mu_\theta^{(t+1)}
\nabla_{\bm\theta}
\mathcal J
(
\bm\theta^{(t)},
\mathbf r^{(t)}
),
\label{eq:theta-update}
\end{equation}
where
$\mu_\theta^{(t+1)}$
is the step size determined by Armijo backtracking line search.
Then, the range parameters are updated as
\begin{equation}
\mathbf r^{(t+1)}
=
\mathbf r^{(t)}
-
\mu_r^{(t+1)}
\nabla_{\mathbf r}
\mathcal J(
\bm\theta^{(t+1)},
\mathbf r^{(t)}
),
\label{eq:r-update}
\end{equation}
where
$\mu_r^{(t+1)}$ is also selected by Armijo backtracking line search.
The updated parameters are confined to the feasible angle--range region and path gains are re-estimated according to \eqref{eq:alpha}.  
Moreover, E-CHARM performs noise-aware gain-based path pruning. 
The noise variance required for pruning is obtained from the subarray-level VALSE, given by $\hat{\sigma}_{\rm n}^2
=
\frac{1}{S}\sum_{s=0}^{S-1}\hat{\sigma}_{{\rm n},s}^2$.
Then, paths satisfying
$|\alpha_\ell^{(t+1)}|^2
<
\epsilon_{\rm gain}\hat{\sigma}_{\rm n}^2$
are removed, where $\epsilon_{\rm gain}$ is a predefined pruning threshold
and $\alpha_\ell^{(t+1)}$ denotes the estimated gain of the $\ell$-th path
at the current iteration. 
The refinement procedure continues until convergence or the maximum iteration number $T_{\rm ref}$ is reached.   
The final outputs of E-CHARM are $\left\{
\hat L,
\hat{\boldsymbol\alpha},
\hat{\bm\theta},
\hat{\mathbf r},
\hat{\mathbf h}
\right\}$. 
The overall procedure of E-CHARM is summarized in Algorithm~\ref{alg:ECHARM}.

\vspace{-2mm}

\subsection{Computational Complexity}

 \begin{algorithm}[t]
\caption{E-CHARM Algorithm}
\label{alg:ECHARM}
\small
\begin{algorithmic}[1]

\REQUIRE
$\mathbf y$,
$T_{\rm ref}$,
$\epsilon_{\rm gain}$

\ENSURE
$\hat L$,
$\hat{\boldsymbol\alpha}$,
$\hat{\bm\theta}$,
$\hat{\mathbf r}$,
and $\hat{\mathbf h}$

\STATE Run CHARM in Algorithm~\ref{alg:CHARM} to obtain the initial estimates 
$
\{(\theta_\ell^{(0)},r_\ell^{(0)},\alpha_\ell^{(0)})\}_{\ell=1}^{L^{(0)}}
$ and $\{\hat{\sigma}^2_{{\rm n},s}\}_{s=0}^{S-1}$

\STATE Estimate measurement noise variance
$
\hat{\sigma}_{\rm n}^2
=
\frac{1}{S}\sum_{s=0}^{S-1}\hat{\sigma}_{{\rm n},s}^2
$

\FOR{$t=0,\ldots,T_{\rm ref}-1$}

    \STATE Update
    $\bm\theta^{(t)}\rightarrow\bm\theta^{(t+1)}$
    via \eqref{eq:theta-update}

    \STATE Update
    $\mathbf r^{(t)}\rightarrow\mathbf r^{(t+1)}$
    via \eqref{eq:r-update}

    \STATE Re-estimate
    $\boldsymbol\alpha^{(t+1)}$
    via \eqref{eq:alpha}

    \STATE Remove paths if
    $
    |\alpha_\ell^{(t+1)}|^2
    <
    \epsilon_{\rm gain}\hat{\sigma}_{\rm n}^2
    $

    \STATE Reconstruct the steering matrix using the remaining paths and re-estimate
    $\boldsymbol\alpha^{(t+1)}$
    via \eqref{eq:alpha}

    \IF{convergence criterion is satisfied}
        \STATE \textbf{break}
    \ENDIF

\ENDFOR

\STATE Set
$
\hat{\bm\theta}=\bm\theta^{(t+1)},
\hat{\mathbf r}=\mathbf r^{(t+1)},
\hat{\boldsymbol\alpha}=\boldsymbol\alpha^{(t+1)}
$

\STATE Set $\hat L$ as the dimension of $\hat{\bm\theta}$ and reconstruct
$\hat{\mathbf h}$ 

\RETURN
$\hat L$,
$\hat{\boldsymbol\alpha}$,
$\hat{\bm\theta}$,
$\hat{\mathbf r}$,
and $\hat{\mathbf h}$

\end{algorithmic}
\end{algorithm}

The computational complexity of E-CHARM consists of two modules: the CHARM-based initialization and the refinement. The complexity of CHARM is given in Sec.~\ref{subsec:CHARM-complexity}.
The refinement module alternately updates $\boldsymbol\theta$ and $\mathbf r$ over $T_{\rm ref}$ iterations.
Within each iteration,  $\boldsymbol\theta$ and $\mathbf r$ are updated $T_\theta$ and $T_r$ times, respectively, using gradient-based optimization. 
The complexity of each gradient evaluation is $\mathcal O(M\hat L)$. 
Given $\hat L=\mathcal O(L)$, the overall complexity of E-CHARM can be approximated as
$
\mathcal O\left(
T_{\rm valse}ML^3
+ T_{\rm em}SL^2
+
T_{\rm ref}(T_\theta+T_r)ML
\right).
$
  
\begin{figure}
 \vspace{-3mm}
    \centering
    \includegraphics[width=0.75\linewidth]{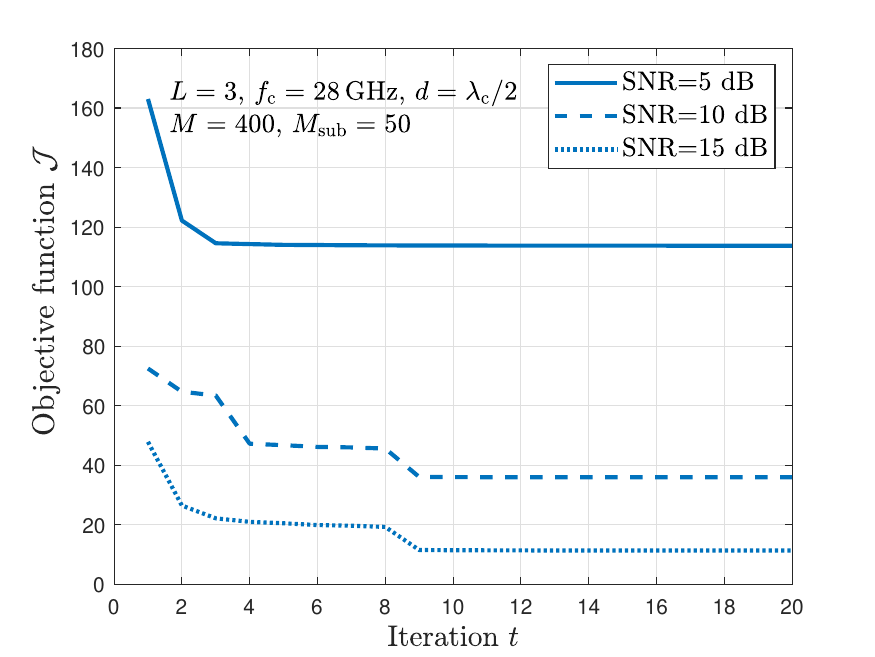}
    \vspace{-1mm}
    \caption{Objective function $\mathcal J$ of E-CHARM at each iteration under different \ac{SNR} levels.}
    \vspace{-2mm}
    \label{fig:ECHARM-cost}
\end{figure}

Note that E-CHARM performs local refinement around the reliable initialization provided by CHARM. 
Therefore, a small number of refinement iterations is sufficient in practice. 
As illustrated in Fig.~\ref{fig:ECHARM-cost}, the objective function converges within a few iterations.
Moreover, the refinement complexity scales linearly with the array size $M$, indicating that E-CHARM introduces moderate additional computational overhead while effectively compensating for the CSF modeling mismatch.

\section{Results and Discussions}\label{sec:simu}

\begin{table*}[th!]
\centering
\caption{Computational Complexity Comparison.}
\label{tab:complexity_compare}
\begin{tabular}{|c|c|c|c|}
\hline
Algorithm & Complexity & Algorithm & Complexity \\
\hline
P-OMP & $\mathcal O(LM^2 N_{\rm range})$ & CHARM & $\mathcal O\Big(
T_{\rm valse}ML^3
+ T_{\rm em}SL^2
\Big)$ \\ \hline
P-SIGW & $\mathcal O(LM^2 N_{\rm range}+T_{\rm sigw}M^2)$ &
E-CHARM & $\mathcal O\Big(
T_{\rm valse}ML^3
+ T_{\rm em}SL^2
+
T_{\rm ref}(T_\theta+T_r)ML
\Big)$ \\
\hline
\end{tabular}
\vspace{-2mm}
\end{table*}

This section evaluates the performance of CHARM and E-CHARM in terms of channel reconstruction, path-number detection, and position estimation accuracy.
For performance comparison, we consider the following benchmarks.

\begin{itemize}

\item \textbf{P-OMP}:
A grid-based polar-domain sparse recovery algorithm~\cite{cui2022channel}. It estimates the near-field channel by iteratively selecting atoms from a polar-domain dictionary with size $M N_{\rm range}$, 
where $M$ and  $N_{\rm range}$ are the numbers of angular and range grid points, respectively.

\item \textbf{P-SIGW}:
An off-grid refinement algorithm initialized by P-OMP~\cite{cui2022channel}. Starting from the coarse polar-domain estimates obtained by P-OMP, it iteratively refines the position parameters under the Fresnel approximation.
The number of refinement iterations is $T_{\rm sigw}$.

\item \textbf{CRLB}:
Assuming the true path number is known, the \ac{CRLB} for position estimation is derived from the parameter-domain Fisher information matrix, while the CRLB for channel estimation is obtained through parameter transformation~\cite{moore2007constrained,kay1993fundamentals}.

\end{itemize}

The computational complexities of the considered algorithms are summarized in Table~\ref{tab:complexity_compare}.  The computational complexities of P-OMP and P-SIGW grow quadratically with the antenna number $M$, while CHARM and E-CHARM achieve linear computational complexity w.r.t. $M$.
 
 \vspace{-2mm}

\subsection{Default Simulation Settings and Performance Metrics}
\label{subsec:simulation-setting}

We consider the carrier frequency $f_{\rm c}=28$ GHz.
The ULA has $M=400$ antennas, which is partitioned into $S=8$ subarrays with $M_{\rm sub}=50$ antennas per subarray. 
Following the simulation setup in \cite{VALSE}, each Monte-Carlo trial contains
$L$ resolvable propagation paths with $L=3$. Specifically, the range and angle of each path are randomly generated within $r_\ell\in[30,40]~{\rm m}$ and
$\theta_\ell\in[-\pi/3,\pi/3]~{\rm rad}$, respectively, subject to the minimum angular and range separations being $\Delta \theta = 0.12~{\rm rad}$ and $\Delta r=1~\rm m$, respectively.
Since the subarray-level angle resolution is $\Delta_{\rm res}=2/M_{\rm sub}=0.04~\rm rad$, these paths are resolvable by each subarray.
The path-gain amplitudes are generated according to $\mathcal N(1,0.1)$, while the phases are independently drawn from $\mathcal U(-\pi,\pi)$.  
The total number of Monte-Carlo trials is $N_{\rm mc}=100$.

For CHARM, we set $L_{\rm valse}=2L$ and $T_{\rm em}=20$.
The thresholds for merging redundant trajectories are set based on the subarray angle resolution as $\epsilon_{\omega}=2\pi d\Delta_{\rm res}/\lambda_{\rm c}$ and
$\epsilon_{\phi}=\pi d^2(1-\Delta_{\rm res}^2)/(\lambda_{\rm c} r_{\max})$.
The minimum association probability for path pruning is $\epsilon_{\rm asso}=0.05$. 
The number of refinement iterations in E-CHARM is $T_{\rm ref}=20$ and weak paths are pruned according to the relative-gain threshold $\epsilon_{\rm gain}=0.01$. 
The default setting of P-OMP and P-SIGW follows
Ref. \cite{cui2022channel}. Since these two methods do not perform
path-number estimation, they are run with $\hat L=2L$, and output the $L$ strongest recovered paths.

The channel reconstruction accuracy is measured by the \ac{NMSE} of the channel estimate, defined as ${\rm NMSE}=
\mathbb{E}\left[\frac{
\|\hat{\mathbf h}-\mathbf h\|_2^2
}{
\|\mathbf h\|_2^2}\right]$.
The path-number detection accuracy is evaluated by the correct detection probability $P(\hat L=L)$.
The position estimation accuracy is only measured over trials with correct path-number detection. The estimated components are matched to the true ones using the Hungarian algorithm~\cite{munkres1957algorithms} with the cost being the \ac{RMSE} of position estimates. 
The RMSE is defined as ${\rm RMSE} 
=
\sqrt{
\mathbb E\left[
\frac{1}{L}
\sum_{\ell=1}^{L}
\left\|
\hat{\mathbf p}_{\ell}
-
\mathbf p_{\ell}
\right\|_2^2
\right]
}$, 
where
$\mathbf p_{\ell}=[r_\ell\cos\theta_\ell,\, r_\ell\sin\theta_\ell]^{\mathsf T}$
and
$\hat{\mathbf p}_{\ell}=[\hat r_\ell\cos\hat\theta_\ell,\,
\hat r_\ell\sin\hat\theta_\ell]^{\mathsf T}$. 
\vspace{-2mm}

\subsection{Simulation Results}

\subsubsection{Performance Under Resolvable Multipath Channel}\label{subsec:simu-basic}
 
\begin{figure*}[th!]
\vspace{-3mm}
    \subfloat[Channel reconstruction accuracy.\label{fig:simu-SNR-NMSE}]
    {\begin{minipage}{.325\textwidth}
        \centering
        \includegraphics[width=1\linewidth]{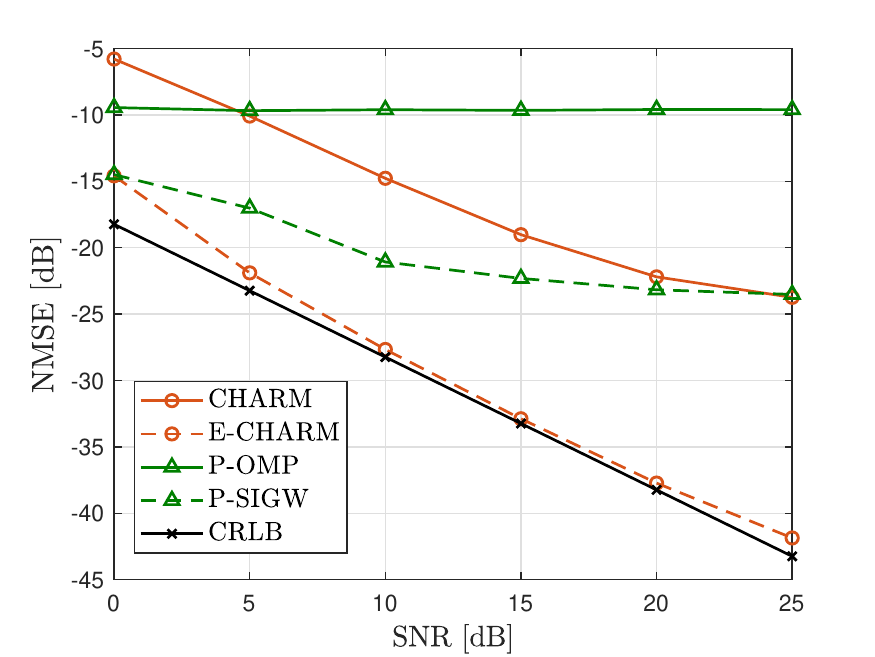}
     \end{minipage}}
    \hfill
    \subfloat[Path-number detection accuracy.\label{fig:simu-SNR-Detection}]
    {\begin{minipage}{.325\textwidth}
        \centering
        \includegraphics[width=1\linewidth]{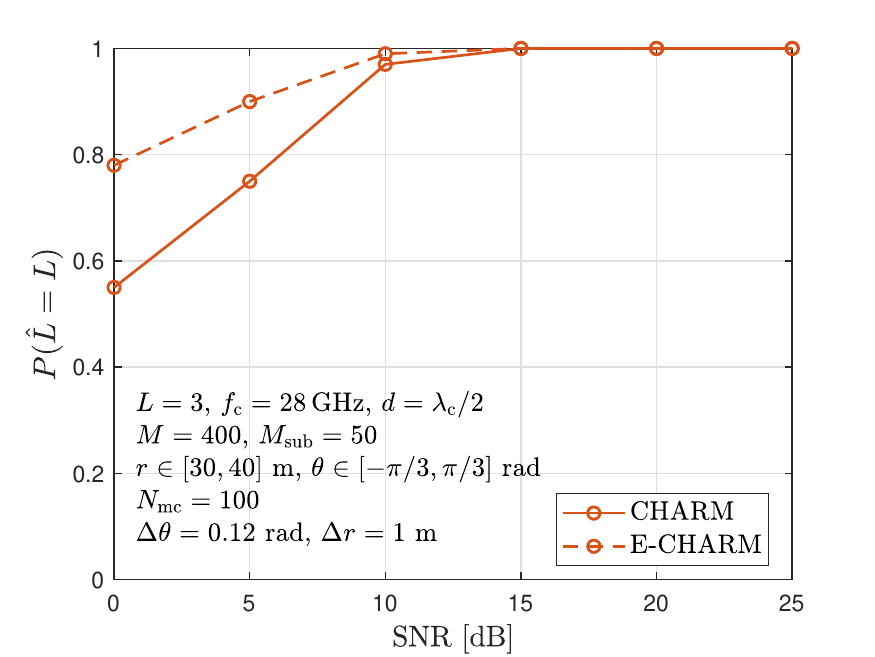}
     \end{minipage}}
    \hfill
    \subfloat[Position estimation accuracy.\label{fig:simu-SNR-RMSE}]
    {\begin{minipage}{.325\textwidth}
        \centering
        \includegraphics[width=1\linewidth]{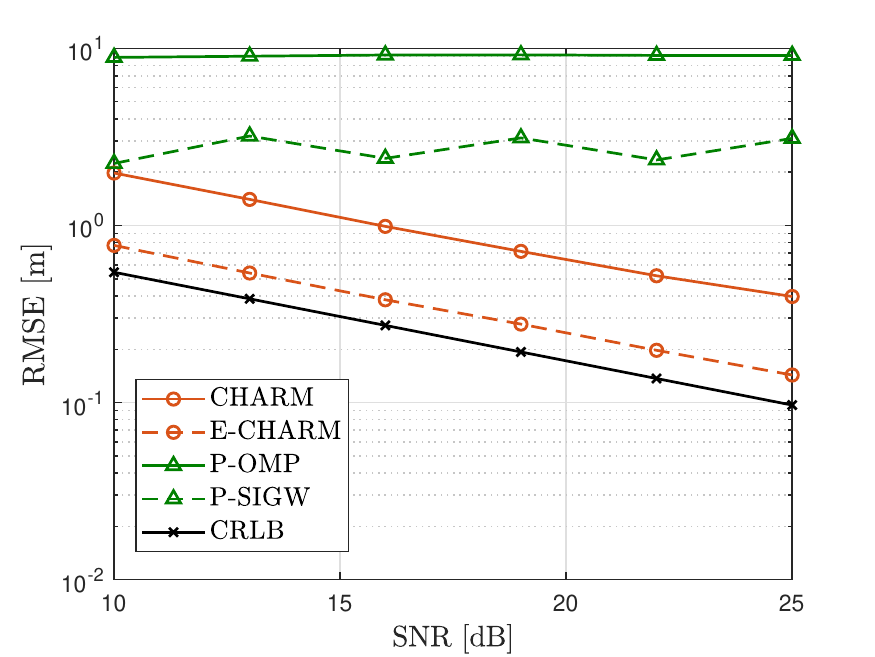}
     \end{minipage}}
     \vspace{-1mm}
    \caption{Performance vs SNR under default simulation settings.}
    \label{fig:simu-SNR}
    \vspace{-3mm}
\end{figure*}

Fig.~\ref{fig:simu-SNR} compares the performance of the considered algorithms under different \ac{SNR} levels.
As shown in Fig.~\ref{fig:simu-SNR-NMSE}, for the channel reconstruction, P-OMP exhibits an almost SNR-independent NMSE floor, indicating that its reconstruction error is dominated by the grid mismatch rather than measurement noise. P-SIGW improves upon P-OMP, but its performance remains constrained by the grid-based initialization. 
By contrast, the NMSE of CHARM decreases steadily with SNR, since more accurate subarray-level VALSE estimates and fewer spurious local components lead to more reliable global fusion.
Starting from the gridless estimate provided by CHARM, E-CHARM further refines the estimate and 
its NMSE closely approaches the CRLB when $\text{SNR}\ge 5~\rm dB$.
The gap between P-SIGW and E-CHARM highlights the importance of reliable initialization for subsequent refinement.

Fig.~\ref{fig:simu-SNR-Detection}  shows that CHARM can reliably infer the unknown path number through subarray-level VALSE and EM-based path association. E-CHARM further improves the detection probability in the low-SNR regime through path-gain-based pruning. Both algorithms achieve a successful detection probability above 95\% when $\text{SNR}\ge 10~\rm dB$.

Conditioned on successful path-number detection,
Fig.~\ref{fig:simu-SNR-RMSE} further evaluates the position-estimation RMSE. Both CHARM and E-CHARM substantially outperform P-OMP and P-SIGW. Interestingly, a comparison with Fig.~\ref{fig:simu-SNR-NMSE} shows that, although P-SIGW achieves a lower channel NMSE than CHARM over most of the considered SNR range, the position RMSE of P-SIGW remains considerably higher. This indicates that accurate fitting of the composite channel, i.e., the superposition of multiple path components, does not necessarily imply faithful recovery of the underlying physical paths, since biased or incorrectly resolved components may still provide an accurate equivalent channel representation.

Table~\ref{tab:runtime_accuracy} further summarizes the runtime and estimation accuracy at $\mathrm{SNR}=20$~dB. 
CHARM achieves the lowest runtime, while E-CHARM substantially improves the estimation accuracy with only a modest additional computational cost. 
Overall, these results demonstrate that CHARM and E-CHARM enable computationally efficient and high-precision near-field channel parameter estimation.

\begin{table}[t]
    \centering
    \caption{Runtime, position RMSE, and channel NMSE under the default
    simulation settings at $\mathrm{SNR}=20$~dB.}
    \label{tab:runtime_accuracy}
    \small
    \setlength{\tabcolsep}{4pt}
    \renewcommand{\arraystretch}{1.15}
    \begin{tabular}{lccccc}
        \toprule
        Metric
        & CHARM & E-CHARM & P-OMP & P-SIGW & CRLB\\
        \midrule
        Runtime [s]
        & 0.225 & 0.266 & 0.803 & 2.612 & \textemdash \\
        RMSE [m]
        & 0.639 & 0.246 & 9.241 & 3.126 & 0.172 \\
        NMSE [dB]
        & -22.1 & -37.7 & -9.5 & -23.1 & -38.2 \\
        \bottomrule
    \end{tabular}  
\end{table}

\subsubsection{Impact of Path Separation}\label{subsec:simu-angle-sep}
 
\begin{figure*}[th!]
\vspace{-3mm}
    \subfloat[Channel reconstruction accuracy.\label{fig:simu-angle-NMSE}]
    {\begin{minipage}{.325\textwidth}
        \centering
        \includegraphics[width=1\linewidth]{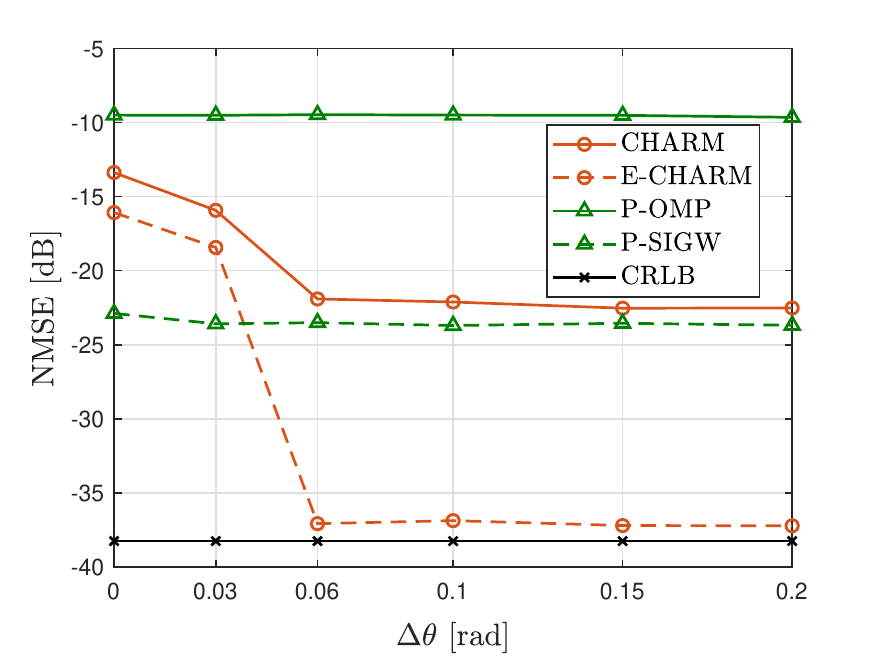}
     \end{minipage}}
    \hfill
    \subfloat[Path-number detection accuracy.\label{fig:simu-angle-Detection}]
    {\begin{minipage}{.325\textwidth}
        \centering
        \includegraphics[width=1\linewidth]{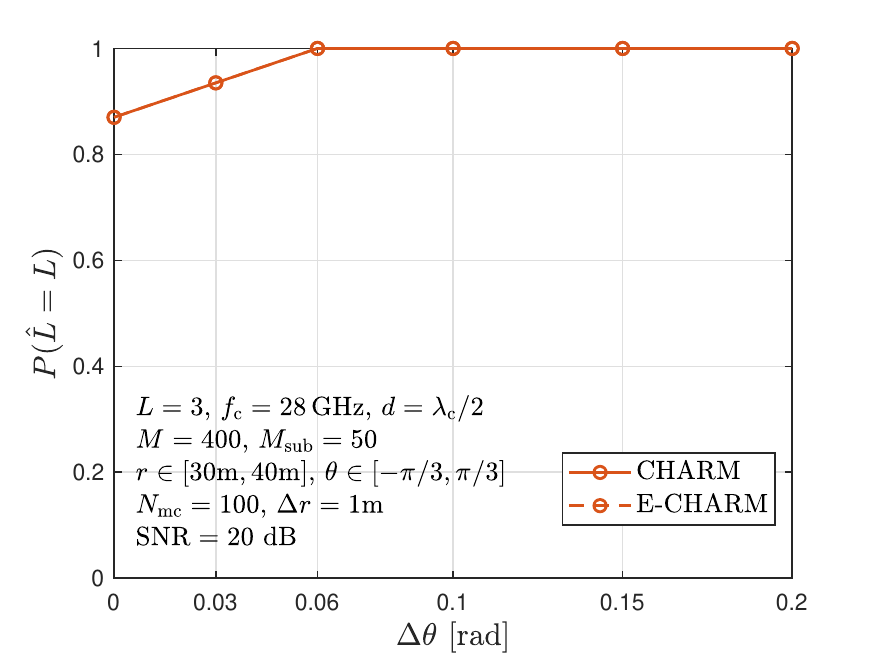}
     \end{minipage}}
    \hfill
    \subfloat[Position estimation accuracy.\label{fig:simu-angle-RMSE}]
    {\begin{minipage}{.325\textwidth}
        \centering
        \includegraphics[width=1\linewidth]{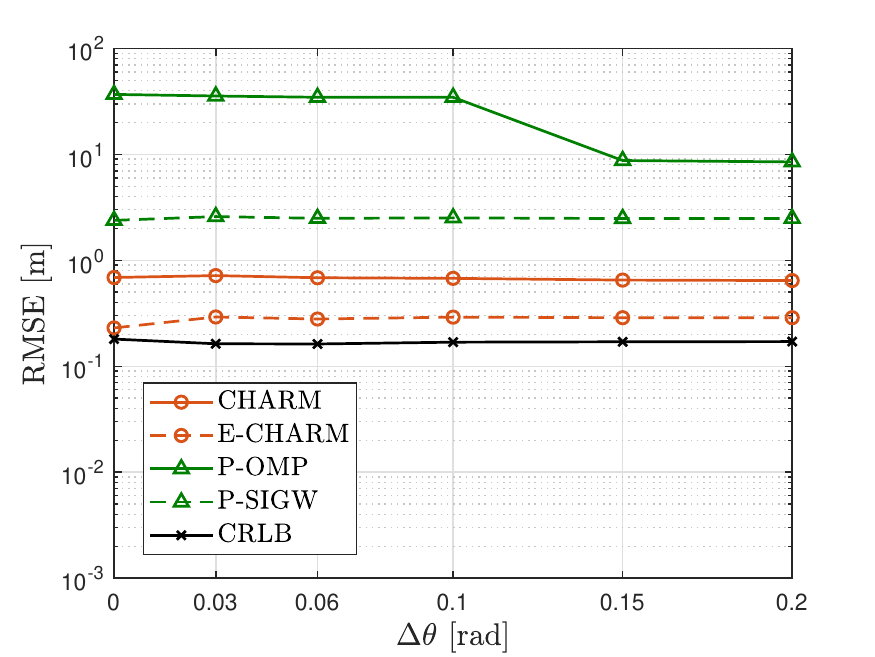}
     \end{minipage}}
     \vspace{-1mm}
    \caption{Performance vs minimum angle separation $\Delta{\theta}$ at $\rm SNR=20~dB$.}
    \label{fig:simu-angle}
    \vspace{-3mm}
\end{figure*}

Fig.~\ref{fig:simu-angle} further investigates the impact of path separation by varying the minimum angle separation $\Delta{\theta}$ at a fixed SNR of $20~\rm dB$, while maintaining the minimum range separation at $\Delta{r}=1~\rm m$.

Figs.~\ref{fig:simu-angle-NMSE} and~\ref{fig:simu-angle-Detection} show that the channel NMSE and path-number detection are strongly governed by the subarray-level angle resolution, i.e., $\Delta_{\rm res}=0.04~\rm rad$.
For $\Delta\theta\le0.03~\rm rad$, closely spaced paths become increasingly difficult to distinguish during subarray-level inference. 
Consequently, path-number recovery becomes less reliable, 
accompanied by a noticeable degradation in the channel NMSEs of CHARM and E-CHARM.
As $\Delta{\theta}$ increases from $0.03$ to $0.06~\rm rad$, the paths become locally resolvable for each subarray, substantially improving the performance of CHARM and E-CHARM. 
When $\Delta\theta\ge0.06~\rm rad$, further increasing the angular separation yields little additional performance improvement, indicating that path proximity is no longer a dominant error source once the paths are locally resolvable.

Conditioned on successful path-number detection,
Fig.~\ref{fig:simu-angle-RMSE} shows that the position RMSEs of CHARM and E-CHARM are nearly insensitive to $\Delta{\theta}$.
Together with Figs.~\ref{fig:simu-angle-NMSE} and~\ref{fig:simu-angle-Detection}, this indicates that reducing the angular separation mainly affects whether all paths can be correctly resolved.
Once successful detection is achieved, both proposed algorithms provide stable position estimates.

Overall, these results suggest an important tradeoff in selecting the subarray size for the proposed algorithms. A larger $M_{\rm sub}$ improves local angle resolution and facilitates the separation of closely spaced paths, but may weaken the validity of the subarray far-field approximation. Conversely, a smaller $M_{\rm sub}$ better preserves the local far-field model at the cost of reduced local angular resolvability.


\subsubsection{Impact of Clustered Scattering}

\begin{figure*}[th!]
\vspace{-5mm}
    \subfloat[Channel reconstruction accuracy.\label{fig:simu-cluster-NMSE}]
    {\begin{minipage}{.325\textwidth}
        \centering
        \includegraphics[width=1\linewidth]{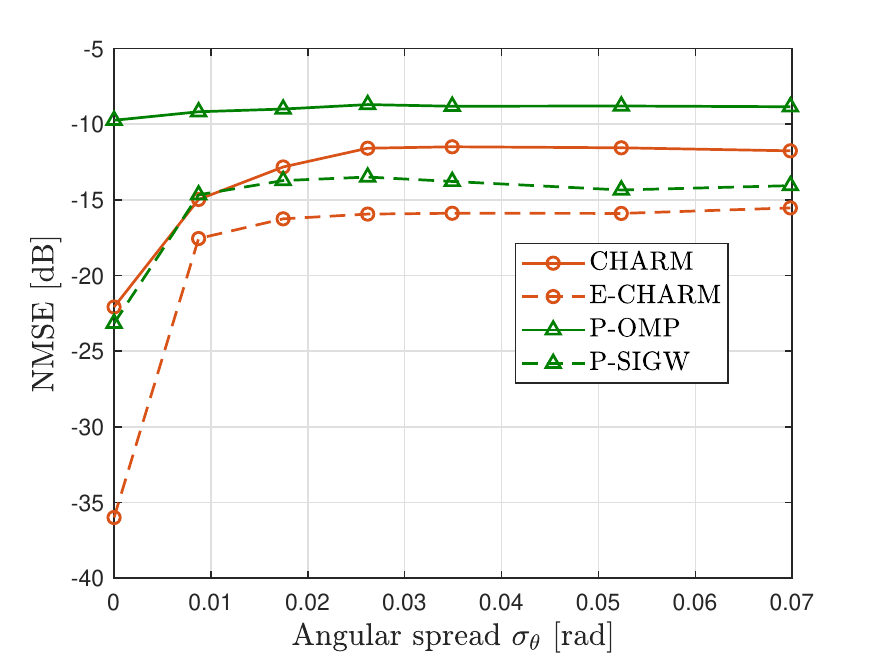}
     \end{minipage}}
    \hfill
    \subfloat[Detection accuracy of dominant path number.\label{fig:simu-cluster-Detection}]
    {\begin{minipage}{.325\textwidth}
        \centering
        \includegraphics[width=1\linewidth]{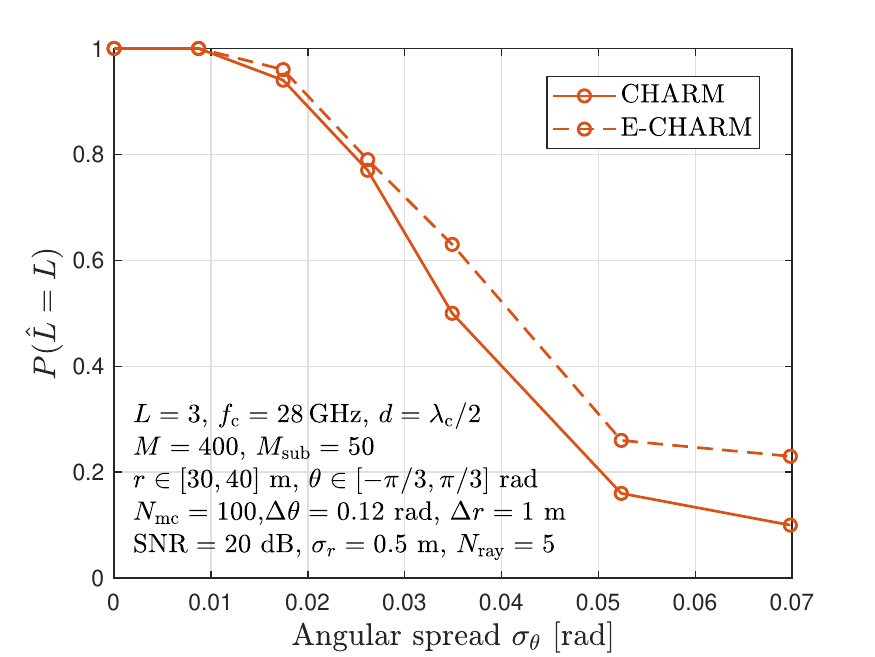}
     \end{minipage}}
    \hfill
    \subfloat[Localization accuracy of dominant scatterers.\label{fig:simu-cluster-RMSE}]
    {\begin{minipage}{.325\textwidth}
        \centering
        \includegraphics[width=1\linewidth]{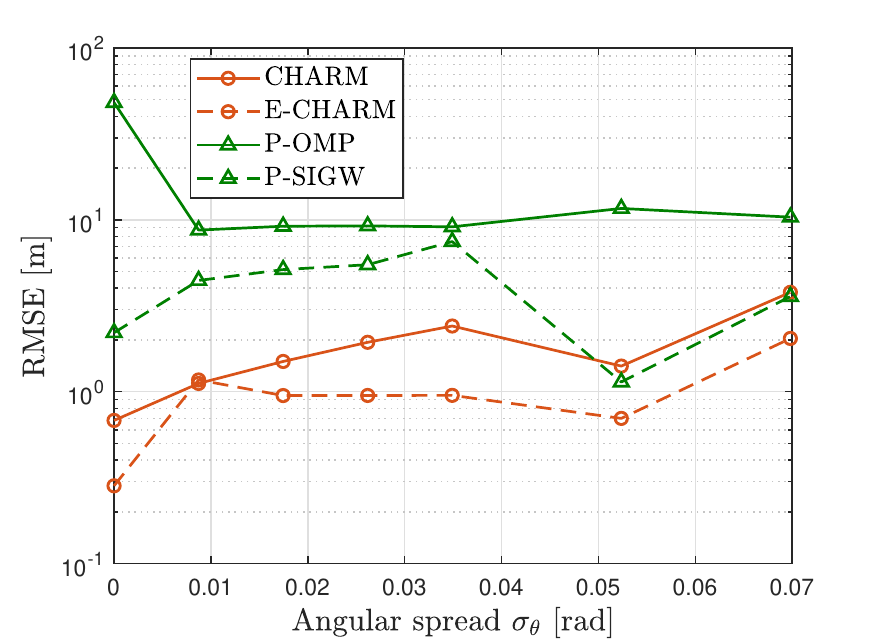}
     \end{minipage}}
     \vspace{-1mm}
    \caption{Performance vs angular spread $\sigma_{\theta}$ under clustered multipath channel at $\rm SNR=20~dB$.}
    \label{fig:simu-cluster}
    \vspace{-3mm}
\end{figure*}

To evaluate the performance of the considered algorithms under more complicated multipath propagation, we further consider a clustered multipath channel comprising $L=3$ scattering clusters. 
Each cluster contains one
dominant scattering component located at $[r_\ell\cos\theta_\ell,r_\ell\sin\theta_\ell]^{\mathsf T}$ and
$N_{\rm ray}=5$ weaker components distributed around it. 
The dominant components are generated according to the default settings in Sec.~\ref{subsec:simulation-setting}. Within each cluster, the range and angle offsets of the weak components follow zero-mean Gaussian distributions with standard deviations $\sigma_r$ and $\sigma_\theta$,
respectively, while  
their power is $15~\rm dB$ below that of the dominant component.
In Fig.~\ref{fig:simu-cluster}, the range spread is fixed at
$\sigma_r=0.5~\rm m$, whereas the angular spread is
varied over $\sigma_\theta \in \{0,0.5,1,1.5,2,3,4\}^{\circ}$, corresponding to $\{0, 0.0087, 0.0175, 0.026, 0.0349, 0.0524, 0.0698\}~\rm rad$. 
The considered algorithms are applied without modification to extract the dominant components and then use them to approximate the clustered channel.

Fig.~\ref{fig:simu-cluster-NMSE} evaluates the effect of clustered scattering on channel reconstruction.
At $\sigma_\theta=0~\rm rad$, favorable performance of the proposed algorithms is still achieved despite the
range spread $\sigma_r=0.5~\rm m$, indicating that CHARM and E-CHARM can tolerate moderate intra-cluster range dispersion.
This is because the weak components within each cluster remain concentrated in angle and therefore produce similar local spatial frequencies over each subarray. 
In contrast, when $\sigma_\theta$ is increased from $0$ to $0.0087~\rm rad$, a nonzero angular spread causes pronounced performance degradation. 
Each cluster exhibits multiple angles due to the weak components. Therefore, angular offsets directly disperse the local spatial frequencies observed by the subarrays, degrading both local inference and global fusion.
As $\sigma_\theta$ further increases, however, the NMSEs of CHARM and E-CHARM remain nearly unchanged. Since the average power of the weak components is fixed, a wider angular spread mainly redistributes this power over a broader angular region without 
substantially increasing its overall contribution to the reconstruction error. 

Fig.~\ref{fig:simu-cluster-Detection} further shows that the probabilities of correctly recovering the dominant-path number decrease with $\sigma_\theta$. As the weak components become more widely dispersed in angle, they increasingly interfere with subarray-level component extraction and cross-subarray association, making it more difficult to identify exactly $L$ dominant paths.
In particular, some weak components may be resolved as additional paths, potentially leading to overestimation of the dominant-path number.
Conditioned on successful dominant-path-number detection, Fig.~\ref{fig:simu-cluster-RMSE} evaluates the localization accuracy of the dominant scatterers, where E-CHARM consistently achieves the lowest position RMSE throughout the considered angular-spread range.

Overall, these results show that E-CHARM provides better robustness to clustered scattering than the benchmark methods, while also revealing the limitation of representing a clustered multipath channel using only its dominant components.

 \section{Conclusion}\label{sec:conclusion}

This paper investigated scalable, high-precision near-field channel parameter estimation using a single uplink observation with unknown path number, based on the spatial chirp structure and array partitioning.
Specifically, based on the spatial chirp structure of the near-field multipath channel, 
we established the CSF model, which connects the local sinusoidal representations within individual subarrays through path-specific linear frequency trajectories. 
This transformed the original high-dimensional channel estimation problem into low-dimensional subarray-level frequency inference and global trajectory recovery. 
We then proposed CHARM, which combines gridless subarray-level line spectral estimation with EM-based cross-subarray association to recover the path number, position parameters, and channel from unordered and imperfect subarray-level estimates.
We further developed E-CHARM to refine the CHARM estimates under the near-field model following the ML principle. 
Both algorithms maintain low computational complexity that scales linearly with the antenna number.
Numerical results confirmed that the proposed algorithms enable high-precision channel parameter estimation and channel reconstruction.

\appendices 

\vspace{-2mm}

\section{Proof of Proposition~\ref{prop:CSF}}\label{app:CSF}

Substituting \eqref{eq:Delta_m} into  
\eqref{eq:phase-near} yields
\begin{align}\label{eq:phase-near2}
    \!\psi_{m,\ell}^{\rm near} 
    \!=\! \tilde{m}_s \omega_\ell + \!\omega_\ell \Delta \tilde{m} + \!\tilde{m}_s^2\phi_\ell
+\!
2 \tilde{m}_s\phi_\ell \Delta \tilde{m}
+\!
\phi_\ell \Delta \tilde{m}^2\!\!.\!\!
\end{align} 
The last term in \eqref{eq:phase-near2}, i.e.,
$\phi_\ell\Delta\tilde m^2$, represents the residual quadratic phase variation within the $s$-th subarray.
Since $|\Delta\tilde m|\le (M_{\rm sub}-1)/2$, we have
\begin{equation}
\begin{aligned}
\max_{\Delta\tilde m}
\left|
\phi_\ell\Delta\tilde m^2
\right|
\le
\frac{\pi d^2}{\lambda_{\rm c} r_\ell}
\left(
\frac{M_{\rm sub}-1}{2}
\right)^2 =
\frac{\pi D_{\rm sub}^2}
{4\lambda_{\rm c}r_\ell},
\end{aligned} 
\end{equation}
where $D_{\rm sub}=(M_{\rm sub}-1)d$ is the subarray aperture.
Following the widely adopted $\pi/8$ phase-error criterion~\cite{AntennaTheory}, this residual quadratic term is negligible if $r_\ell
\ge
\frac{2D_{\rm sub}^2}{\lambda_{\rm c}}$,
which coincides with the classical Rayleigh-distance criterion associated with the subarray aperture~\cite{selvan2017fraunhofer}.
In this case, 
$\psi_{m,\ell}^{\rm near}$ can be approximated as
\begin{equation}\label{eq:phase-subarray}
\psi_{m,\ell}^{\rm near}=
\tilde{m}_s\omega_{\ell}+\tilde{m}_s^2\phi_\ell+ 
\Delta \tilde{m} ( \omega_\ell + 2\phi_\ell \tilde{m}_s ), 
\end{equation}
where $\tilde{m}_s\omega_{\ell}+\tilde{m}_s^2\phi_\ell$ is invariant within the $s$-th subarray and can be regarded as a common phase shift, while $\Delta \tilde{m} ( \omega_\ell + 2\phi_\ell \tilde{m}_s )$ denotes the phase difference across the subarray antennas.
As a result, the phase variation within each subarray can be approximated as a linear function of $\Delta\tilde{m}$. 
Based on \eqref{eq:chirp-model} and \eqref{eq:phase-subarray}, the received signal can be approximated as \eqref{CSFmodel}.

\vspace{-2mm}

\section{Proof of Proposition~\ref{prop:local_angle}}
\label{app:subarrayAoA}

Consider a source point located at $[r_\ell\cos\theta_\ell, r_\ell\sin\theta_\ell]^\mathsf T$,
where $\theta_\ell$ is the global angle observed at the array center.
The center of the $s$-th subarray is located at $[0,\tilde m_s d]^\mathsf T$, which observes the local angle $\theta_{s,\ell}$. 
Let $r_{s,\ell}$ denote the distance between the $\ell$-th source point and the center of the $s$-th subarray. 
From Fig.~\ref{fig:SubarrayAoA}, the global angle $\theta_{\ell}$ satisfies
\begin{equation}
\begin{aligned}
r_{s,\ell}^2
&=
(r_\ell\cos\theta_\ell)^2
+
(r_\ell\sin\theta_\ell-\tilde m_s d)^2  \\
&=
r_\ell^2
+
(\tilde m_s d)^2
-
2r_\ell\tilde m_s d\sin\theta_\ell .
\end{aligned}
\label{eq:rs_exact}
\end{equation} 
Moreover, the local angle $\theta_{s,\ell}$ satisfies
\begin{equation}
\sin\theta_{s,\ell}
=
\frac{r_\ell\sin\theta_\ell-\tilde m_s d}{r_{s,\ell}}. 
\label{eq:theta_s_exact}
\end{equation}

To obtain an intuitive relation between $\theta_{s,\ell}$ and $\theta_\ell$,
we first rewrite \eqref{eq:rs_exact} as  
\begin{align}
\frac{r_{s,\ell}}{r_\ell}
&\!= \!
\sqrt{
1
+
(t_s^2
-
2t_s\sin\theta_\ell)
}  \overset{(a)}{=}
1
-
t_s\sin\theta_\ell
+ \mathcal O(t_s^2)
,\!\!
\label{eq:rs_first_order}
\end{align}
where $t_s
\triangleq
\frac{\tilde m_s d}{r_\ell}$, and (a) applies the first-order Taylor expansion $\sqrt{1+x}=1+\frac{x}{2}+\mathcal O(x^2)$ with $x=t_s^2
-
2t_s\sin\theta_\ell$. 
Substituting \eqref{eq:rs_first_order} into \eqref{eq:theta_s_exact}, we have
\begin{equation}
\sin\theta_{s,\ell}
=
\frac{\sin\theta_\ell-t_s}
{r_{s,\ell}/r_\ell}
=\frac{\sin\theta_\ell-t_s}
{1-t_s\sin\theta_\ell
+\mathcal O(t_s^2)}.
\label{eq:sin_theta_ratio_approx}
\end{equation}
Next, applying the first-order reciprocal expansion
$\frac{1}{1-x}
=
1+x+\mathcal O(x^2)$  with $x=t_s\sin\theta_\ell+\mathcal O(t_s^2)$, we obtain
\begin{align}
\sin\theta_{s,\ell}
&=
(\sin\theta_\ell-t_s)
(1+t_s\sin\theta_\ell)
+
\mathcal O(t_s^2)
\nonumber\\ 
&=
\sin\theta_\ell
-
t_s\cos^2\theta_\ell
+
\mathcal O(t_s^2).
\label{eq:sin_theta_s_expanded}
\end{align}
Retaining the first-order term in \eqref{eq:sin_theta_s_expanded} \ac{w.r.t.} $t_s$ gives \eqref{eq:LocalAngle}.

\bibliographystyle{IEEEtran}
\bibliography{reference}
\end{document}